\documentclass{aa}
\usepackage{amsmath, amssymb, amsfonts, amscd}
\usepackage[varg]{txfonts} % math symbols, as bold Times (next amsmath)

\usepackage{graphicx} % to include graphics
\usepackage{epstopdf}

\usepackage{natbib}
\bibpunct{(}{)}{;}{a}{}{,} % to follow the A&A style

\usepackage[pdffitwindow=false,plainpages=false,bookmarksopen=false,breaklinks=true,pdfpagelabels=true,bookmarksnumbered=true,colorlinks=true, linkcolor=blue, citecolor=blue]{hyperref} 

\begin{document}

\title{Tidal inertial waves in the differentially rotating convective envelopes of low-mass stars}
\subtitle{I. Free oscillation modes}

\author{M. Guenel\inst{1}
\and C. Baruteau\inst{2}
\and S. Mathis\inst{1,3}
\and M. Rieutord\inst{2}
}

\institute{Laboratoire AIM Paris-Saclay, CEA/DSM - CNRS - Universit\'e Paris Diderot, IRFU/SAp Centre de Saclay, F-91191 Gif-sur-Yvette Cedex, France
\and Institut de Recherche en Astrophysique et Plan\'etologie, Observatoire Midi-Pyr\'en\'ees, Universit\'e de Toulouse, 14 avenue Edouard Belin, 31400 Toulouse, France
\and LESIA, Observatoire de Paris, PSL Research University, CNRS, Sorbonne Universit\'es, UPMC Univ. Paris 6, Univ. Paris-Diderot, Sorbonne Paris Cité, 5 place Jules Janssen, 92195 Meudon, France\\
\email{[mathieu.guenel,stephane.mathis]@cea.fr;[clement.baruteau,michel.rieutord]@irap.omp.eu} 
}

\date{Received ... / accepted ...}

%=ABSTRACT==%=========%=========%=========%=========%=========%=========%=
\abstract
% Contexte
{Star-planet tidal interactions may result in the excitation of inertial waves in the convective region of stars. In low-mass stars, their dissipation plays a prominent role in the long-term orbital evolution of short-period planets. Turbulent convection can sustain differential rotation in their envelope, with an equatorial acceleration (as in the Sun) or deceleration, which can modify the waves' propagation properties.}
% Aim
{We explore in this first paper the general propagation properties of free linear inertial waves in a differentially rotating homogeneous fluid inside a spherical shell. We assume that the angular velocity background flow depends on the latitudinal coordinate only, close to what is expected in the external convective envelope of low-mass stars.}
% Method
{We use i) an analytical approach in the inviscid case to get the dispersion relation, from which we compute the characteristic trajectories along which energy propagates. This allows us to study the existence of attractor cycles and infer the different families of inertial modes ; ii) high-resolution numerical calculations based on a spectral method for the viscous problem.}
% Results
{We find that modes that propagate in the whole shell (D modes) behave the same way as with solid-body rotation. However, another family of inertial modes exists (DT modes), which can propagate only in a restricted part of the convective zone. Our study shows that they are less common than D modes and that the characteristic rays and shear layers often focus towards a wedge -- or point-like attractor. More importantly, we find that for non-axisymmetric oscillation modes, shear layers may cross a corotation resonance with a local accumulation of kinetic energy. Their damping rate scales very differently from what we obtain for standard D modes and we show an example where it is independent of viscosity (Ekman number) in the astrophysical regime in which it is small.}
% Conclusions
{}

%=====KEY-WORDS=====%=========%=========%=========%=========%=========%=

\titlerunning{Tidal inertial waves in the differentially rotating convective envelopes of low-mass stars - I}
\authorrunning{Guenel et al.}
%\authorrunning{Guenel, Baruteau, Mathis, Rieutord}

\maketitle

%=========%=========%=========%=========%=========%=========%=========%=
% TEXT
%=========%=========%=========%=========%=========%=========%=========%=

\section{Introduction}
\label{sec:introduction}

The tidal interaction between a star and an orbiting companion results from the differential force exerted by each body on the other. Indeed, the gravitational potential created by the companion is not uniform inside the star since it is an extended body and not a point-like mass. If the orbit is close enough, the star experiences a non-negligible periodical forcing which generates tidal flows. Their dissipation into heat through various friction mechanisms \citep[see][]{Zahn2013, MathisRemus2013, Ogilvie2014} takes energy away from the system and allows for a redistribution of angular momentum, which leads to the evolution of the spin and orbital parameters on secular timescales. Depending on the initial distribution of angular momentum between the individual spins and the orbit, the system may evolve towards an equilibrium state --- where the orbit is circular and the spins of the bodies are synchronized and aligned with the orbit --- or lead to the inspiral or ejection of the companion \citep{Hut1980}.

Observational evidence for ongoing tidal interactions was found in close binary stars, for instance by \cite{Giuricin1984}, \cite{Verbunt1995} and more recently, \cite{Meibom2005} who showed that the oldest systems have circularized orbits, though the spin-orbit synchronization is still unclear \citep{Meibom2006}. Observations of extrasolar planets by the radial-velocity and transits methods have developed rapidly over the past decade and stimulated interest in looking for signatures of tidal interactions in star-planet systems : for instance, \cite{Pont2009} looked for an excess rotation in stars with planets (due to substantial inward migration) while \cite{Husnoo2012} used observed eccentricities to conclude that tidal interactions play a prominent role in the orbital evolution and survival of hot Jupiters \citep[see also][]{Lai2012, Guillot2014}. Moreover, \cite{Jackson2008} consistently checked that the observed low eccentricities of close-in planets can be explained by tidal evolution processes, and calibrated the required tidal quality factors. \cite{Hansen2010} and \cite{Penev2012} used observed populations of exoplanets and tidal evolution models to constrain tidal dissipation in stars, and both studies seem to agree that their results are inconsistent with the dissipation inferred from the circularization of binary stars, suggesting that a different mechanism is at play in this case. \cite{Winn2010} and \cite{Albrecht2012} also proposed that tidal dissipation in convective regions could be responsible for the low obliquities of hot Jupiters orbiting around cool stars, though \cite{Mazeh2015} found that this property extends to Kepler Objects of Interest with orbital periods up to at least 50 days, for which tidal interactions should be negligible. These results have to be put into perspective since many other processes are responsible for shaping the architecture of the systems --- such as migration in the protoplanetary disk \citep[e.g.][]{Baruteau2014}, planet-planet scattering \citep[e.g.][]{Chatterjee2008}, Kozai oscillations \citep[e.g.][]{Naoz2011}, star-planet magnetic interactions \citep[see][]{Strugarek2014}, magnetic spin-down of the star \citep[see][]{Barker2009,Damiani2015}, etc. --- and the global picture remains quite uncertain.

On the theoretical side, the tidal response of a star has been investigated in the past, starting with the theory of the equilibrium tide, which is the flow induced by the quasi-hydrostatic adjustment to the tidal potential, and its dissipation by turbulent friction in the convective envelope of low-mass stars \citep{Zahn1966_a, Zahn1966_b}. This work was followed by the study of the radiative damping of tidal gravity modes \citep{Zahn1975} in more massive stars with an outer radiative region. \cite{Zahn1977} then estimated the strengths of these mechanisms in order to derive the characteristic timescales for circularization, synchronization and alignment for binary stars. Since then, most works about tidal interactions were based on Zahn's approach and focused on giving more realistic descriptions of the mechanisms thought to be responsible for tidal dissipation --- especially the dynamical tide in radiative regions in early-type stars. For instance, \cite{Savonije1983, Savonije1984} performed numerical simulations of gravity modes excited by a periodic tidal potential in a non-rotating massive star and derived the associated timescales for the orbital evolution of massive binaries. The effects of rotation on these gravity modes and the associated tidal dissipation, ignored by Zahn, were then investigated by \cite{Rocca1987, Rocca1989} who included the effects of the Coriolis acceleration, and \cite{Goldreich1989} who took into account possible internal differential rotation.

After the discovery of the first exoplanetary systems in the mid 1990's, the study of star-planet tidal interaction regained interest and studies shifted towards lower-mass stars with radiative cores and convective envelopes, for which the accepted tidal dissipation mechanism so far was the viscous dissipation of Zahn's equilibrium tide in the convective zone. \cite{Terquem1998}, \cite{Goodman1998} and \cite{Witte2002} performed numerical calculations of the dynamical tide in the radiative core of non-rotating solar-like stars and studied the effect of resonances in the core of a solar-type star, and found that the associated tidal dissipation could be more efficient than the viscous dissipation of the equilibrium tide in the convective envelope. \cite{Barker2010, BarkerOgilvie2011, Barker2011} also studied the possible non-linear breaking of tidal gravity waves near the centre of a solar-type star and its influence on tidal dissipation and on the survival of hot Jupiters and angular momentum deposition. Progressively, the low-frequency regime --- where rotation is likely to have the most important effects --- of stellar oscillations regained interest with the works by \cite{Savonije1995, Savonije1997, Papaloizou1997} who simulated the full tidal response of a uniformly-rotating massive star and found signatures of tidally-excited inertial waves in the convective core, as well as gravito-inertial waves in the radiative envelope. \cite{Witte1999, Witte2001} studied the tidal evolution of massive binary systems including the effects of resonant excitation of gravito-inertial modes and possible resonance locking, which can speed up significantly the secular evolution.

Unlike stably stratified radiative zones, stellar convective regions are very slightly unstably stratified and gravity modes cannot propagate inside them. For more than thirty years, the possible effects of a dynamical tide in convective zones have been overlooked in favour of Zahn's equilibrium tide theory. However, a simple approach to model this dynamical tide is to simply ignore the convective motions and to consider linear disturbances to a neutrally stratified fluid. In this context, the dynamical tide consists of low-frequency oscillations primarily restored by the Coriolis acceleration --- otherwise known as inertial waves --- that propagate in a sphere (if the body is fully convective) or a spherical shell (for sun-like stars with an outer convective envelope). These waves have been studied long ago in a uniformly rotating, incompressible and inviscid fluid by \cite{Thomson1880, Poincare1885, Bryan1889, Cartan1922} (also \cite{Greenspan1968}) and they exhibit remarkable properties : their frequency $\tilde{\omega}$ in the fluid frame (rotating with angular velocity $\Omega$) is restricted to $[-2\Omega, 2\Omega]$ while their spatial structure is governed by a hyperbolic second-order partial differential equation named after Poincar\'e. They propagate along rays that are inclined with a constant angle $\arcsin\left(\tilde{\omega}/2\Omega\right)$. Such hyperbolic equations along with boundary conditions in a closed container generally yield an ill-posed problem whose solutions may be singular. The behavior of inertial waves is driven by the rays properties : in a full sphere, they fill the whole domain for all frequencies and a set of global normal modes with frequencies that is dense in $[-2\Omega,2\Omega]$ has been found in the case of a full sphere \citep{Bryan1889}. This is not true in the case of a spherical shell \citep{Rieutord1997}, for which rays often focus towards limit cycles called attractors \citep[see][]{Maas1995, Rieutord2001} that exist in narrow frequency bands and depend on the size of the inner core. The inclusion of viscosity --- as a simplification of turbulence --- gives rise to peculiar modes that possess shear layers localized around the aforementioned attractors \citep[e.g.][]{Rieutord2001}.

The tidal force exerted by a close planetary or stellar companion on its host star may therefore excite inertial waves in the external convective envelope of low-mass stars, if one or more components of the tidal potential has a frequency that fall into the inertial range. The energy dissipated and the angular momentum carried by these waves may play an important role on the evolution of the orbital architecture of close-in planetary systems and on the rotation of their components, yet it has only recently started to be investigated \citep[see][]{OgilvieLin2004, OgilvieLin2007, Ogilvie2009, Goodman2009, Rieutord2010, LeBars2015} in uniformly-rotating fluids. The tidal dissipation induced by forced inertial modes has been found to be a very erratic function of the excitation frequency and varies over several orders of magnitude, though frequency-averaged estimates can be obtained analytically \citep{Ogilvie2013} and evaluated along stellar evolution \citep{Mathis2015}, showing strong variations with stellar mass, age and rotation. It has been shown by \cite{BR2013} that differential rotation may strongly affect the propagation properties of linear inertial modes of oscillations. Their study was restricted to shellular and cylindrical rotation profiles, but turbulent convection can also establish various differential rotation profiles \citep[see][]{Brun2002, Brown2008, Matt2011,Gastine2014}, including conical rotation as observed in the Sun \citep{Schou1998, Garcia2007}. In this work, we explore the propagation and dissipation properties of inertial modes in stellar convective envelopes (for low-mass stars only) with conical differential rotation, namely differential rotation that only depends on latitude.

In Sect. \ref{sec:setup}, we expose our physical model and the associated hypotheses as well as the differential rotation profile we use. Then in Sect. \ref{sec:inviscid_problem}, we carry out an analytical analysis which yields the general propagation properties of inertial waves in the inviscid case. These results are compared to the viscous problem in Sect. \ref{sec:viscous_problem} where we compute numerically the velocity fields of such inertial modes, which sometimes have very different properties than in the solid-body rotation case. Finally, we identify in Sect. \ref{sec:conclusion} which of these properties are important for the forced problem that we will treat in a subsequent paper, and in particular the evaluation of the tidal dissipation induced by these waves.

\section{Inertial modes in differentially rotating convective envelopes}
\label{sec:setup}

\subsection{Physical model}
\label{sec:physical_model}

Since we want to study the propagation of inertial waves in the differentially rotating convective envelopes of low-mass-stars, we will consider a simplified setup with an incompressible, viscous fluid inside a spherical shell of external radius $R$ and internal radius $\eta R$ ($0 < \eta < 1$) that accounts for the boundary between the radiative core and the convective envelope.

As motivated in Sect. \ref{sec:introduction}, we use a conical differential rotation profile --- depending only on the colatitude $\theta$ --- which reads

\begin{equation}
\Omega(\theta) = \Omega_0(\theta) / \Omega_{\rm ref} = 1 + \varepsilon \sin^2\theta,
\label{eq:rotation_profile}
\end{equation}
so that the dimensionless angular velocity of the background flow is $1$ at the poles and $1+\varepsilon$ at the equator. The quantity $\varepsilon$ is a parameter that describes the behavior of the differential rotation :

\begin{itemize}
\item $\varepsilon > 0$ is for solar differential rotation (equatorial acceleration),
\item $\varepsilon < 0$ is for anti-solar differential rotation (equatorial deceleration).
\end{itemize}
The value $\epsilon \approx 0.3$ approximates the differential rotation profile in the Sun's convective envelope.

We point out that the above rotation profile does not satisfy the Taylor-Proudman theorem, which states that the velocity field of a rotating incompressible homogeneous inviscid fluid must be constant along columns parallel to the rotation axis. As is now well-known \cite[e.g.][]{Brun2002,Brown2008}, this differential rotation profile is sustained by Reynolds stresses due to convective turbulent motions. This mean flow is assumed to be dynamically stable. Our analysis will show that some modes are unstable for some parameters of the model (see Eq. (\ref{eq:dispersion_general})). These instabilities, interesting {\it per se}, should be interpreted as giving the limits of the parameters range of this model, especially in its future use for tidally forced flows.

The waves that we study in the following are low-frequency waves: their frequency is less than $2\Omega_{\rm max}$, where $\Omega_{\rm max}$ is the maximum angular velocity of the fluid. Such waves propagate on a turbulent background and presumably interact with eddies that have similar time scale (just like the stochastically excited acoustic modes of solar-type stars \citep{Zahn1966_b, Goldreich1977}). In a Kolmogorov type turbulence, the turn-over time scale of eddies decreases as $\ell^{2/3}$ if $\ell$ is the scale of the eddies. The equality between the turnover time scale and rotation period determines what we call the ``Rossby scale" below which turbulence is little influenced by rotation and is fast compared
to the wave period. Our model assumes that eddies smaller than the Rossby scale can be represented by a turbulent viscosity. To fix
ideas, let us consider the example of the Sun. The bounding frequency $2\Omega_{\rm max}$ corresponds to waves with a period of 12.5 days. From Mixing-Length Theory, which says that the typical scale of turbulence is the mixing -length (about twice the pressure scale height), we note that the largest eddies have a typical scale of 100Mm, with a typical velocity of 50~m/s. This would be the typical numbers for the so-called giant cells characterized by a 20 days turn over time scale \citep{Miesch2008}. With these numbers we easily find that the Rossby scale is 50Mm. Thus all the eddies of scale much smaller than 50Mm are assumed to influence the inertial modes through a turbulent diffusion. The influence of turbulent motions with larger scales is admittedly much more difficult to take into account. However, we may note that in the case of the Sun the differential rotation flow is much stronger than the large scale eddies (1~km/s compared to 50~m/s). Thus, as a first step, neglecting the direct interaction of these eddies with the inertial waves is consistent as far as orders of magnitude are concerned.

\subsection{Mathematical formulation}

In an inertial frame with the usual spherical coordinates $(r,\theta,\varphi)$, and after linearization of the Navier-Stokes equation around the steady state (described by the ``0'' subscripts), we obtain \citep[e.g.][]{BR2013}
\begin{equation}
\label{eq:linearized_NS}
\frac{\partial {\bf u_1}}{\partial t} + \Omega_0 \frac{\partial {\bf u_1}}{\partial \varphi} + 2 \Omega_0 {\bf e_z} \times {\bf u_1} + r \sin \theta \left({\bf u_1} \cdot \nabla\Omega_0 \right) {\bf e_{\varphi}} = -\nabla p_1 + \nu \nabla^{2}{\bf u_1} ,
\end{equation}
where $\partial/\partial t$ denotes the partial time-derivative, $\Omega_0$ is the background angular velocity of the fluid, ${\bf e_z} = \cos\theta{\bf e_r} -\sin\theta{\bf e_{\theta}}$ is the unit vector along the rotation axis, $\vec{\bf u_1}$ is the wave velocity field and $p_1$ denotes the reduced pressure perturbation, which is the pressure perturbation divided by the reference density ($\rho_0$). Note that only the last term of the left-hand side is proportional to the shear of the differential rotation, while the other terms are formally identical to the case of solid-body rotation.

Along with this equation, we use the linearized continuity equation
\begin{equation}
\label{eq:continuity}
\nabla \cdot {\bf u_1} = 0
\end{equation}
since the fluid is also assumed to be incompressible.

We look for velocity ($\bf u_1$) and reduced pressure ($p_1$) perturbations of angular frequency $\Omega_p$ (in the inertial frame) and azimuthal wavenumber $m$. This means they are proportional to $\exp\left(i\Omega_p t + im\varphi\right)$. Eqs. (\ref{eq:linearized_NS}) and (\ref{eq:continuity}) can be recast as
\begin{equation}
\begin{cases}
& i \tilde{\Omega}_p {\bf u_1} + 2 \Omega_0 {\bf e_z} \times {\bf u_1} + r \sin\theta \left( {\bf u_1}\cdot\nabla\Omega_0 \right) {\bf e_{\varphi}} = -{\nabla}p_1 + \nu \nabla^2 {\bf u_1},\\
& \nabla \cdot {\bf u_1} = 0,
\end{cases}
\label{eq:system}
\end{equation}
where $\tilde{\Omega}_p = \Omega_p+m\Omega_0$ is the Doppler-shifted wave frequency (that is the local wave frequency in a frame rotating with the fluid).

We normalize all frequencies by $\Omega_{\rm ref}$ --- which we define as the background angular velocity at the poles \emph{i.e.} $\Omega_0(\theta=0)$ --- and distances by $R$. We introduce the associated dimensionless quantities : the radius $x=r/R$, the velocity field ${\bf u} = {\bf u_1} / R \Omega_{\rm ref}$, the reduced pressure $p = p_1/R^2 \Omega_{\rm ref}^2$, the differential rotation profile $\Omega = \Omega_0/\Omega_{\rm ref}$, the wave frequency $\omega_p = \Omega_p / \Omega_{\rm ref}$ in the inertial frame (resp. $\tilde{\omega}_p = \tilde{\Omega}_p / \Omega_{\rm ref}$ in the fluid's frame) and the Ekman number $E = \nu/ {R^2 \Omega_{\rm ref}}$. From now on, we shall use exclusively the dimensionless quantities defined above. This finally yields the dimensionless system that we shall solve numerically in Sect. \ref{sec:viscous_problem} :
\begin{equation}
\begin{cases}
& i \tilde{\omega}_p {\bf u} + 2 \Omega {\bf e_z} \times {\bf u} + x \sin\theta \left( {\bf u}\cdot\nabla\Omega \right) {\bf e_{\varphi}} = -\nabla p + E \nabla^2 {\bf u}, \\
& \nabla \cdot {\bf u} = 0.
\end{cases}
\label{eq:dimensionless_system}
\end{equation}

Note that in the numerical simulations presented in Sect. \ref{sec:viscous_problem} below, we actually take the curl of the first equation in order to get rid of the $\nabla p$ term :

\begin{equation}
\begin{cases}
& \nabla \times \left( i \tilde{\omega}_p {\bf u} + 2 \Omega {\bf e_z} \times {\bf u} + x \sin\theta \left( {\bf u}\cdot\nabla\Omega \right) {\bf e_{\varphi}} \right) =  E \nabla \times \nabla^2 {\bf u}, \\
& \nabla \cdot {\bf u} = 0.
\end{cases}
\label{eq:viscous_problem}
\end{equation}

In addition to Eqs. (\ref{eq:viscous_problem}), we use stress-free boundary conditions (${\bf u} \cdot {\bf e_r} = 0$ and ${\bf e_r} \times [\sigma]{\bf e_r} = \bf 0$, where $[\sigma]$ is the viscous stress tensor) at the inner and outer boundaries of the spherical shell.

In the following section, we define several useful quantities for the description of the propagation properties of inertial waves, and Eq. (\ref{eq:rotation_profile}) implies that they only depend on the colatitude $\theta$ and are independent of the aspect ratio $\eta$ of the spherical shell.

\section{Inviscid analysis : paths of characteristics, existence of turning surfaces and corotation resonances}
\label{sec:inviscid_problem}

As a first step, we study the solutions to Eqs. (\ref{eq:dimensionless_system}) in the inviscid limit ($E = 0$). It allows us to study the dynamics of characteristics of inertial waves, which are a very useful tool for understanding the solutions to the full viscous problem for small viscosities.

\subsection{The general Poincar\'e-like equation}

Following \cite{BR2013}, we adopt cylindrical coordinates $(s,\varphi,z)$ and we eliminate the components of the velocity perturbations, which yields a partial differential equation on $p$ only. Focusing on the second-order terms, we obtain the following mixed-type partial differential equation :
\begin{equation}
\label{eq:poincare}
\frac{\partial^2 p}{\partial s^2} + \frac{A_z}{\tilde{\omega}_p^2} \frac{\partial ^2 p}{\partial s \partial z} + \left( 1- \frac{A_s}{\tilde{\omega}_p^2} \right)\frac{\partial^2 p}{\partial z^2} + \dots = 0,
\end{equation}
where zeroth and first order terms have been omitted -- that corresponds to the WKBJ approximation. In Eq. (\ref{eq:poincare}),
\begin{equation}
A_s(s,z) = \frac{2 \Omega}{s} \frac{\partial (s^2 \Omega)}{\partial s} \quad \mbox{and} \quad A_z(s,z) = \frac{2 \Omega}{s} \frac{\partial (s^2 \Omega)}{\partial z}.
\end{equation}

In the case of solid-body rotation ($\varepsilon = 0$), Eq. (\ref{eq:poincare}) reduces to the well-known Poincar\'e equation for inertial waves \citep{Cartan1922, Greenspan1968}.
It is hyperbolic in the domain where the discriminant $\xi(s,z) = A_z^2 + 4 \tilde{\omega}_p^2 \left( A_s - \tilde{\omega}_p^2 \right)$ is positive (which we will refer to as the ``hyperbolic domain'') and becomes elliptic in the domain where $\xi$ is negative (the ``elliptic domain''). In the former case, the equation governing the paths of characteristics in a meridional plane reads
\begin{equation}
\frac{dz}{ds} = \frac{1}{2\tilde{\omega}_p^2} \left( A_z \pm \xi^{1/2} \right),
\label{eq:slope}
\end{equation}
which shows that, contrary to the case of solid-body rotation ($A_z = 0$, $\xi$ constant), differential rotation tends to curve paths of characteristics since the right-hand side of Eq. (\ref{eq:slope}) now depends both on $s$ and $z$. On the other hand, no characteristics exist in the elliptic domain, therefore the relation $\xi = 0$ defines ``turning surfaces'' on which characteristics reflect \citep[see][]{Friedlander1982}. Note that if the rotation profile is symmetric by $z\rightarrow -z$, then Eq. (\ref{eq:poincare}) is also symmetric by $z\rightarrow -z$, meaning only positive values of $z$ can be considered. We detail below the case of the conical rotation profile defined in Eq. (\ref{eq:rotation_profile}).

\subsection{Paths of characteristics, turning surfaces and classes of inertial modes}
\subsubsection{Paths of characteristics}
\label{sec:paths_characteristics}

For the rotation profile that we adopt, given in Eq. (\ref{eq:rotation_profile}), we have
\begin{equation}
A_s = 4 \Omega(\theta) \left(\Omega(\theta) + \varepsilon \sin^2 \theta \cos^2 \theta\right)
\end{equation}
and
\begin{equation}
A_z = -4 \Omega(\theta) \varepsilon \cos \theta \sin^3 \theta,
\end{equation}
so that
\begin{multline}
\xi = 16 \Omega^2(\theta) \varepsilon^2 \cos^2 \theta \sin^6 \theta \\ +4\tilde{\omega}_p^2 \left(4\Omega^2(\theta)+4\Omega(\theta)\varepsilon \cos^2 \theta \sin^2 \theta -\tilde{\omega}_p^2\right).
\end{multline}
We assume that $\varepsilon > -1$ so that the Rayleigh stability criterion ($A_s > 0$ everywhere in the shell) is always satisfied. From Eq. (\ref{eq:slope}), the slope of the paths of characteristics in the hyperbolic domain reads
\begin{equation}
\frac{dz}{ds} = -A(\theta) \cos \theta \sin^3 \theta + {\left(\frac{\xi}{4\tilde{\omega}_p^4}\right)}^{1/2}
\label{eq:slope2}
\end{equation}
where 
\begin{equation}
\frac{\xi}{4\tilde{\omega}_p^4} = A^2(\theta) \cos^2 \theta \sin^6 \theta +2A(\theta)\cos^2 \theta \sin^2 \theta + \frac{4\Omega^2(\theta)-\tilde{\omega}_p^2}{\tilde{\omega}_p^2},
\end{equation}
and
\begin{equation}
A(\theta) = 2 \varepsilon \frac{\Omega(\theta)}{\tilde{\omega}_p^2}.
\end{equation}
As expected, all the above quantities as well as the dynamics of characteristics depend on $\theta$ only.

\subsubsection{Turning surfaces}
\label{sec:turning_surfaces}
 \begin{figure*}
  \centering
  \includegraphics[width=0.4\textwidth]{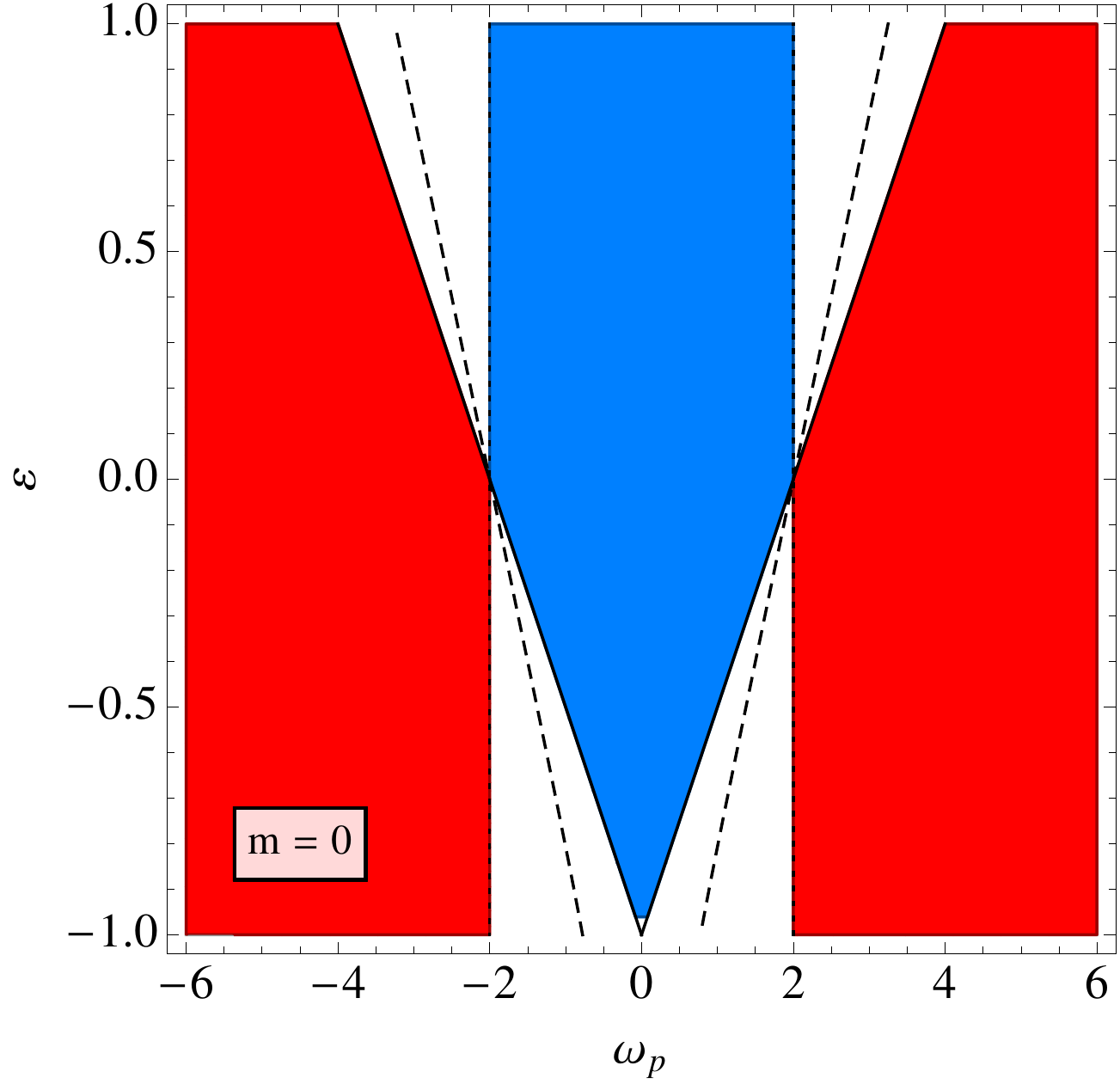} \qquad
  \includegraphics[width=0.4\textwidth]{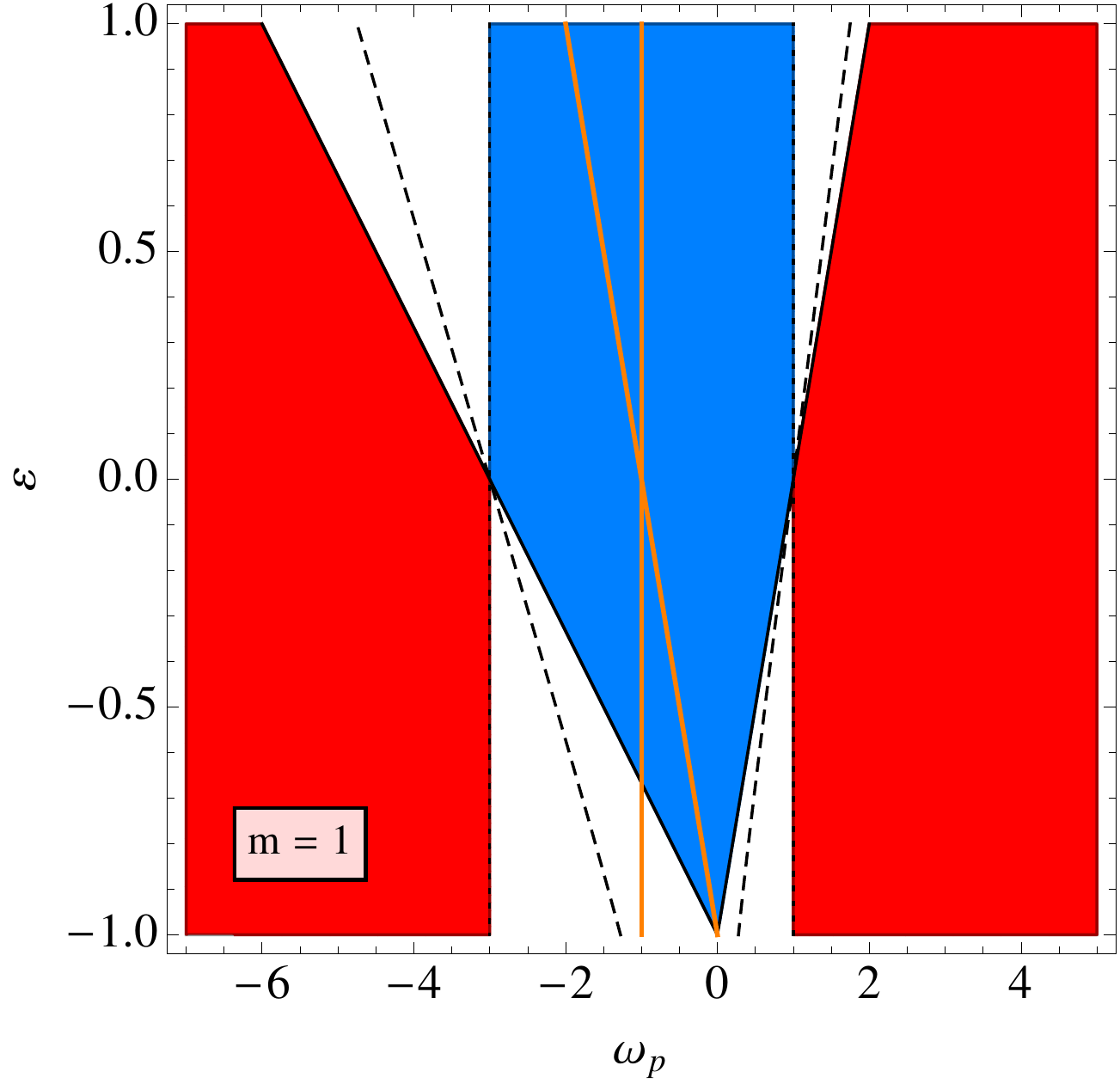}
  \includegraphics[width=0.4\textwidth]{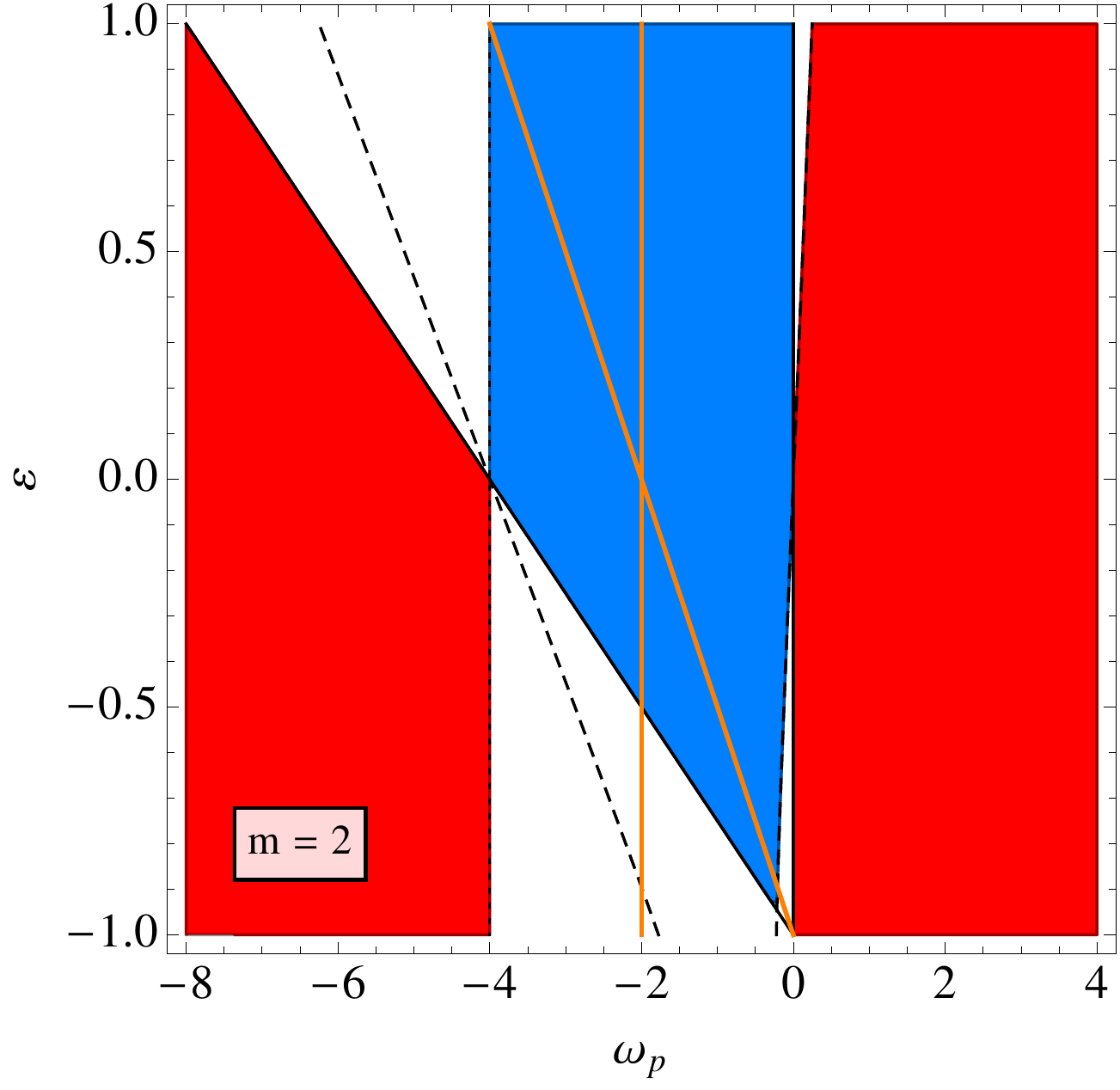} \qquad
  \includegraphics[width=0.4\textwidth]{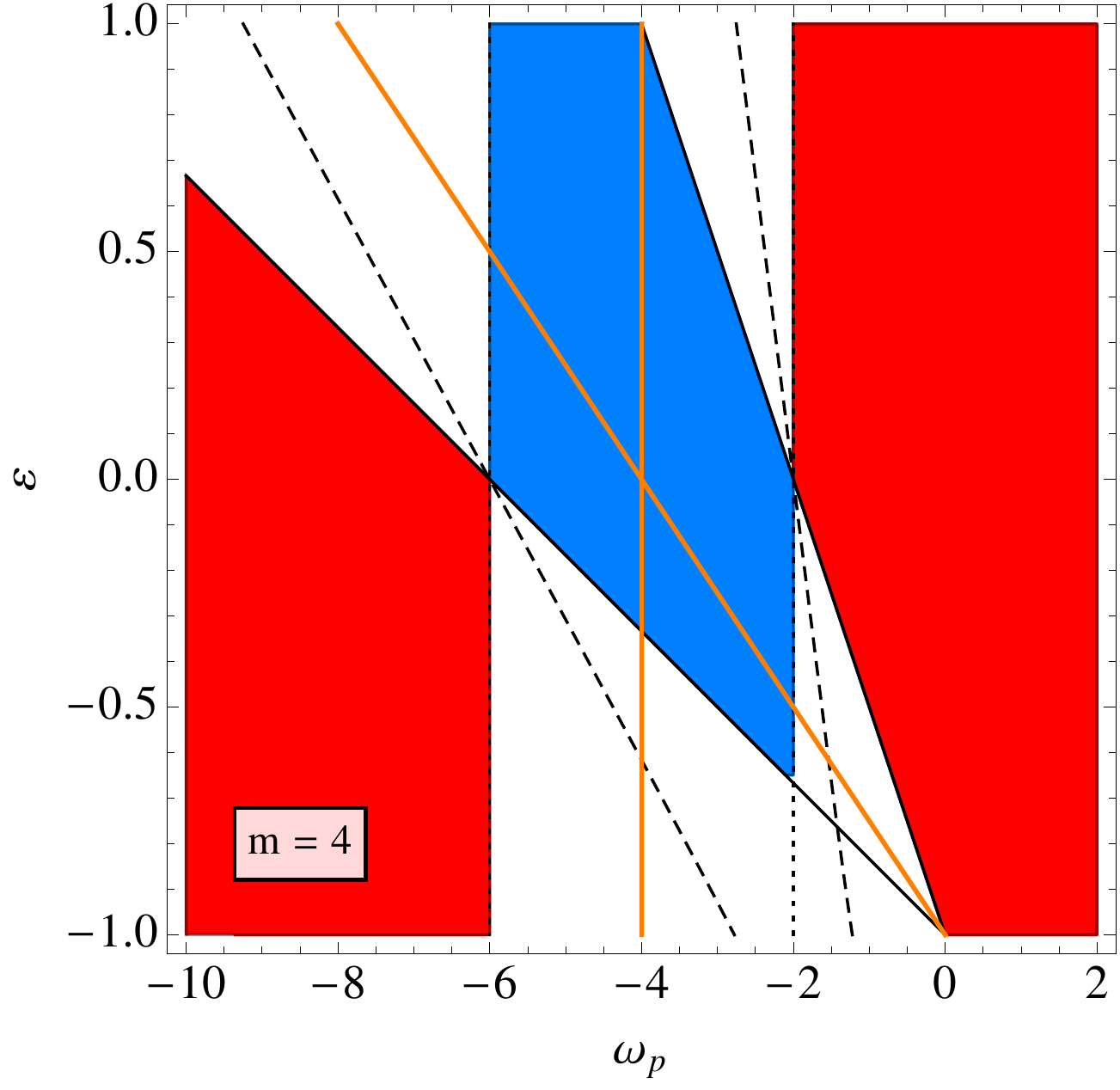}
  \caption{Illustration of the two kinds of inertial modes propagating in a fluid with conical rotation profile $\Omega(\theta) = \Omega_0(\theta)/\Omega_{\rm ref} = 1+\varepsilon \sin^2\theta$, for $m=\{0, 1, 2, 4\}$ from top-left to bottom-right. The eigenfrequency in the inertial frame $\omega_p$ is shown in x-axis and the differential rotation parameter $\varepsilon$ is in y-axis. {\bf Blue~:} D modes that exist in the whole shell ($\xi>0$ everywhere). {\bf White~:} DT modes that exhibit at least one turning surface inside the shell (the sign of $\xi$ changes in the shell). {\bf Red~:} No inertial modes may exist in the shell ($\xi<0$ everywhere). {\bf Orange~:} The modes in the region between the two orange lines feature a corotation layer ($\tilde{\omega}_p = 0$) inside the shell. The meaning of the black solid, dotted and dashed lines is explained in paragraph \ref{sec:turning_surfaces}.}
  \label{fig:BBR}
  \end{figure*}
  
In the setup described in \ref{sec:paths_characteristics}, $\xi$ is symmetric by the equatorial plane of the shell so only values $\theta \in [0,\pi/2]$ need to be considered. As explained in the previous paragraph, equation $\xi(\theta) = 0$ defines turning surfaces that separate the spherical shell into an hyperbolic domain ($\xi >0$) and an elliptic domain ($\xi<0$). Turning surfaces are portions of cones of axis $z$ and aperture $2\theta$. However, their location cannot be determined analytically because $\xi(\theta)=0$ is a polynomial equation of degree 6 in $X=\sin^2 \theta$. Still their existence leads to two classes of inertial modes with conical rotation :
\begin{enumerate}
\item modes for which at least one turning surface exists within the shell, referred to as DT modes following the terminology introduced in \cite{BR2013} (D for differential rotation, T for turning surface),
\item modes that propagate in the whole shell, which is entirely hyperbolic because there is no turning surface within it, named D modes.
\end{enumerate}

The occurrence of D and DT modes is depicted in Fig. \ref{fig:BBR} for the axisymmetric case ($m=0$) and a few non-axisymmetric cases ($m>0$). White areas represent DT modes which have $\xi < 0$ at least once within the shell, while blue regions represent D modes for which $\xi > 0$ everywhere within the shell. The red areas illustrate the case where $\xi < 0$ everywhere in the shell, for which the shell hosts no inertial modes. The curves that separate D and DT modes satisfy :
\begin{itemize}
\item $\xi(\theta=0)=0$, which corresponds to the case where the turning surface is on the rotation axis, and which can be recast as 
\begin{equation}\omega_p = -m \pm 2, \end{equation}
and is shown by the vertical dotted lines in Fig. \ref{fig:BBR},\\
\item $\xi(\theta=\pi/2)=0$, when the turning surface reaches the equator, which yields
\begin{equation}\omega_p = (-m \pm 2)(1+\varepsilon), \end{equation}
and is shown by the solid black lines in Fig. \ref{fig:BBR}.
\end{itemize}

We point out that there may exist two turning surfaces in a meridional plane for a given set of parameters. For instance, in the bottom-left panel of Fig. \ref{fig:BBR} (where $m=2$), the transition from the central blue region to the white region around $\omega_p=0$ shows that two turning surfaces occur at $\theta = \pi/4$ before splitting towards the pole and the equator. Our investigations showed that this only occurs for $m=\pm2$ around $\omega_p=0$, which we checked by solving numerically the equation $\xi(\theta=\pi/4)=0$ for every $m$, whose solutions are depicted by the black dashed lines in Fig. \ref{fig:BBR}. An example is given in the following subsection (see the bottom-right panel of Fig. \ref{fig:attractors2}).

We point out that for D modes --- for which $\forall \theta, \, \xi(\theta)>0$ --- we have $|\tilde{\omega_p}(\theta)|<2\Omega(\theta)$, which means that for D modes, the standard criterion for propagation of inertial waves is met everywhere locally.

\subsubsection{Critical layers}
For non-axisymmetric modes ($m\neq0$), $\tilde{\omega}_p$ may vanish in the shell, which corresponds to so-called corotation resonances or critical layers \citep{BR2013}. Since $\tilde{\omega}_p$ is a function of $\theta$ only, a critical layer is the intersection of a cone of axis $z$ with the spherical shell. As can be seen from Eq. (\ref{eq:slope2}), paths of characteristics may become locally vertical at corotation layers.
Their location is given by :
\begin{equation}
\sin \theta = \sqrt{\varepsilon^{-1}\left( -1-\frac{\omega_p}{m}\right)}
\label{eq:corotation}
\end{equation}
and they exist only when $\omega_p \in \left[-m,-m(1+\varepsilon)\right]$ (if $m\varepsilon < 0$) or $\omega_p \in \left[-m(1+\varepsilon),-m\right]$ (if $m\varepsilon > 0$). Critical layers exist for the D and DT modes located between the two orange lines in the panels of Fig. \ref{fig:BBR}.

\subsection{Dispersion relation, phase and group velocities}

We expand in this paragraph the study of the propagation of inertial waves in the inviscid limit by considering the wave dispersion relation. The latter is obtained by assuming solutions to Eq. (\ref{eq:poincare}) in the form $p \propto \exp[i(k_s s + k_z z)]$ and adopting the short-wavelength approximation :
\begin{equation}
\tilde{\omega}_p^2  = \frac{k_z^2}{\| \bf k \|^2} \left[A_s - \frac{k_s}{k_z}A_z\right],
\label{eq:dispersion_general}
\end{equation}
where $\left\|\bf{k}\right\| = \sqrt{k_s^2+k_z^2}$. Assuming the bracket-term is positive, which is equivalent to assuming that the background rotation profile is dynamically stable to inertial waves (see the discussion in Sec. \ref{sec:physical_model}), we can define :
\begin{equation}
\tilde{\mathcal{B}} = {\left[A_s - \frac{k_s}{k_z}A_z\right]}^{1/2}
\end{equation}
so that the wave dispersion becomes $\tilde{\omega}_p  = \pm \tilde{\mathcal{B}} |k_z| / \| \bf k \|$. The phase and group velocities in a frame rotating with the fluid are then given by :
\begin{equation}
{\bf v_p} = \pm \tilde{\mathcal{B}} \frac{k_z}{\left\|\bf{k}\right\|^3} {\bf k}
\end{equation}
and
\begin{equation}
{\bf v_g} = \pm \frac{k_s}{\left\|\bf{k}\right\|^3} \left(-k_z {\bf e_s} + k_s {\bf e_z} \right) \left[ \tilde{\mathcal{B}} +\frac{A_z}{2} \frac{\left\|\bf{k}\right\|^2}{k_s k_z} \tilde{\mathcal{B}}^{-1} \right],
\end{equation} which corresponds to Eqs. (A9) and (A10) in \cite{BR2013}.
Note that ${\bf k} \cdot {\bf v_g} = 0$ as for pure inertial waves propagating in a solid-body rotating fluid.

Applying these formulae to our conical rotation profile $\Omega(\theta) = 1+\varepsilon \sin^2 \theta$, the dispersion relation is given by
\begin{equation}
\tilde{\omega}_p^2 = 4 \Omega^2(\theta) \frac{k_z^2}{\left\|\bf{k}\right\|^2}\left[ \left(1 + \frac{\varepsilon \sin^2 \theta \cos^2 \theta}{1+\varepsilon \sin^2 \theta}\right) +\frac{k_s}{k_z} \frac{\varepsilon \cos \theta \sin^3 \theta}{1+\varepsilon \sin^2 \theta} \right].
\label{eq:dispersion}
\end{equation}
Assuming the fluid is stable against the GSF instability, we define
\begin{equation}
\mathcal{B} = {\left[ \left(1 + \frac{\varepsilon \sin^2 \theta \cos^2 \theta}{1+\varepsilon \sin^2 \theta}\right) +\frac{k_s}{k_z} \frac{\varepsilon \cos \theta \sin^3 \theta}{1+\varepsilon \sin^2 \theta} \right]}^{1/2},
\end{equation}
so that Eq. (\ref{eq:dispersion}) becomes
\begin{equation}
\tilde{\omega}_p = \pm 2 \Omega(\theta) \mathcal{B} \frac{\left|k_z\right|}{\left\|\bf{k}\right\|}.
\label{eq:dispersion2}
\end{equation}
For solid-body rotation, $\mathcal{B}=1$ and Eq. (\ref{eq:dispersion2}) is the usual dispersion relation for inertial waves. Consequently, the phase velocity in the frame rotating with the fluid is
\begin{equation}
{\bf v_p} = \pm 2 \Omega(\theta) \mathcal{B} \frac{k_z}{\left\|\bf{k}\right\|^3} \bf{k},
\label{eq:phase_velocity}
\end{equation}
while the group velocity is given by
\begin{multline}
{\bf v_g} = \pm 2 \Omega(\theta) \frac{k_s}{\left\|\bf{k}\right\|^3} \left(-k_z {\bf e_s} + k_s {\bf e_z} \right) \\
\times \left[ \mathcal{B} -\frac{2\Omega(\theta) \varepsilon \cos \theta \sin^3 \theta}{k_s k_z} \left\|\bf{k}\right\|^2 \mathcal{B}^{-1} \right].
\label{eq:group_velocity}
\end{multline}
From Eq. (\ref{eq:dispersion2}), we see that a critical layer may form in the shell ($\tilde{\omega_p}=0$) in the following formal cases :
\begin{itemize}
\item $\left|k_s\right| \rightarrow \infty$ at finite $k_z$, so that ${\bf v_p} = \bf 0$ and ${\bf v_g} = \bf 0$ at corotation : inertial waves cannot cross critical layers.
\item $\left|k_z\right| \rightarrow 0$ at finite $k_s$, so that ${\bf v_p} = \bf 0$, ${\bf v_g} \cdot {\bf e_s} = \bf 0$ and $\left|{\bf v_g} \cdot {\bf e_z} \right| \rightarrow \infty$, indicating that inertial waves may propagate across a critical layer with locally vertical paths of characteristics.
\item $\mathcal{B} \rightarrow 0$, which implies again that ${\bf v_p} = \bf 0$ while $\left|{\bf v_g} \cdot {\bf e_s} \right| \rightarrow \infty$ and $\left|{\bf v_g} \cdot {\bf e_z} \right| \rightarrow \infty$, which means that in this case as well inertial waves may propagate across the corotation resonance with a finite slope for the paths of characteristics, since $\mathcal{B}=0$ implies
\begin{equation}
\frac{dz}{ds} = -\frac{k_s}{k_z} = -\frac{1 + \varepsilon \sin^2 \theta \left(1+\cos^2 \theta\right)}{\varepsilon \cos \theta \sin^3 \theta}.
\end{equation}
\end{itemize}

This result shows that when it comes to critical layers, conical and shellular rotation profiles behave similarly --- in the sense that the three aforementioned cases are possible --- while cylindrical rotation stands out as the case where characteristics of inertial waves can formally never cross critical layers \citep[see][]{BR2013}. It is important to keep in mind that this analysis only reflects the hyperbolic part of the problem while the solutions to the full viscous problem (Sect. \ref{sec:viscous_problem}) may behave a bit differently than the characteristics of the purely inviscid problem \citep[as shown for instance by Fig. 13 of][]{BR2013}.

\subsection{Dynamics of the characteristics of inertial waves : attractor cycles, focusing points and Lyapunov exponents}

\begin{figure*}
\centering
\includegraphics[width=0.4\textwidth]{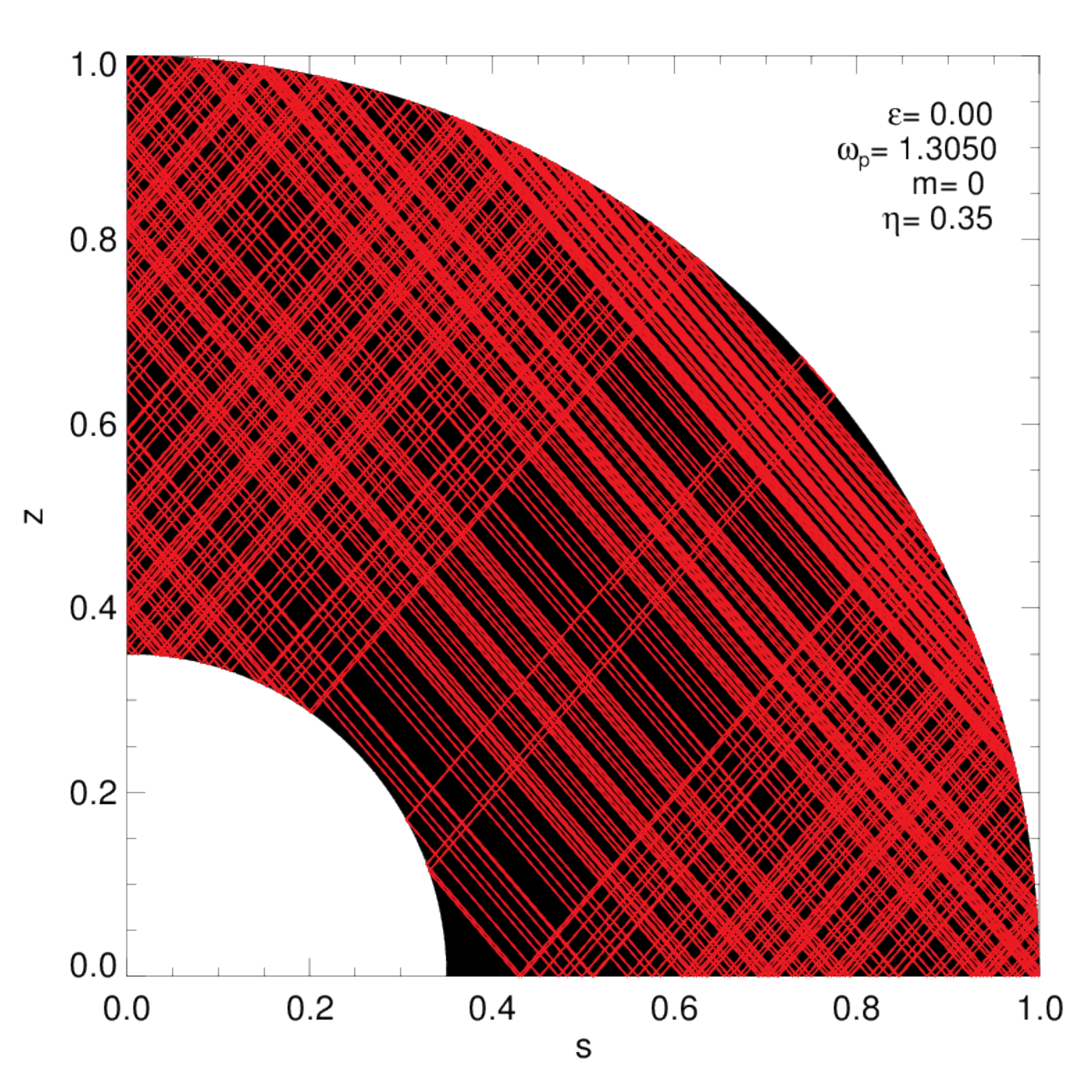} \qquad
\includegraphics[width=0.4\textwidth]{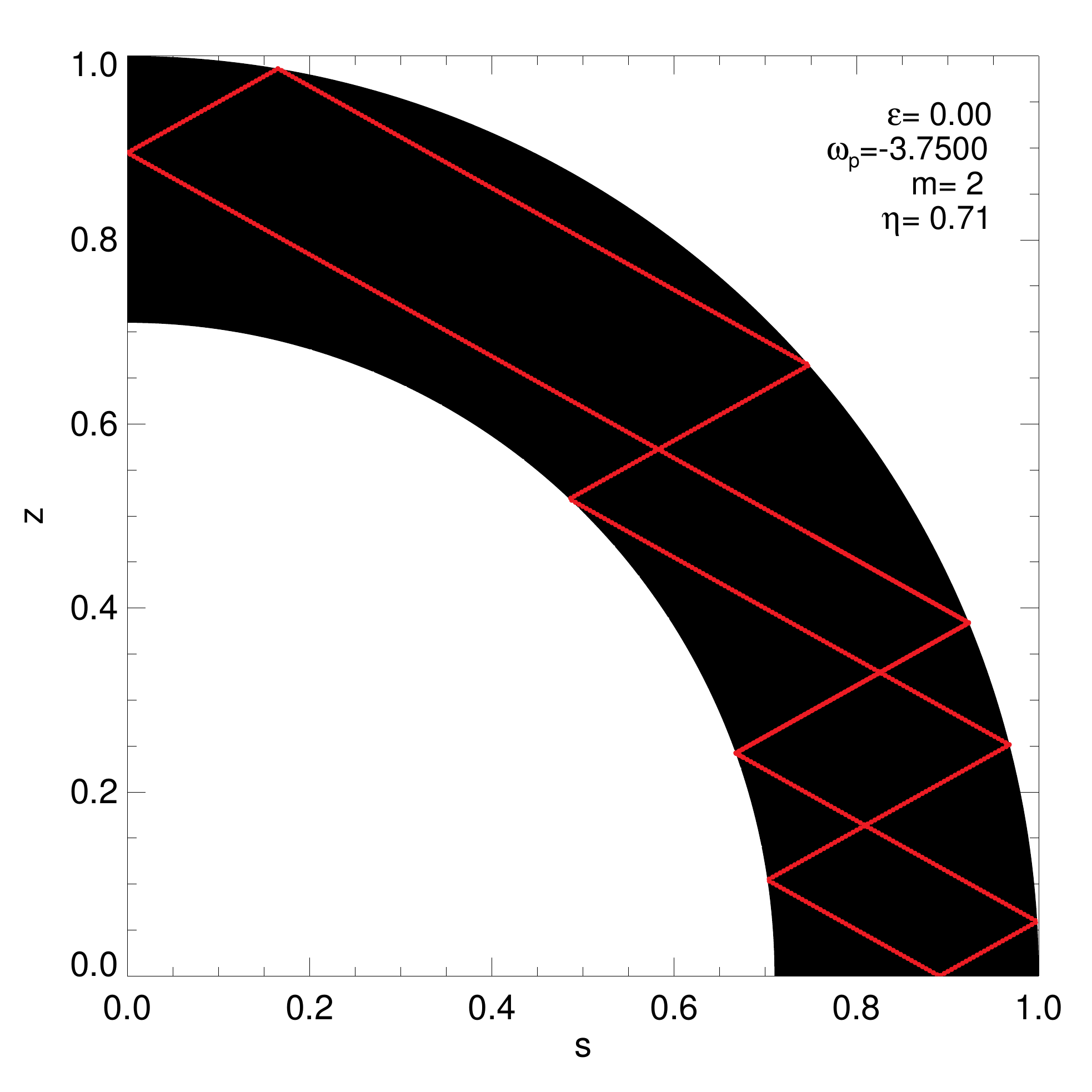}
\caption{Paths of characteristics for solid-body rotation ($\varepsilon = 0$). {\bf Left~:} Paths of characteristics do not converge towards any periodic orbit ($m=0$, $\omega_p=1.305$, $\eta=0.35$ and solid body-rotation $\varepsilon = 0$). {\bf Right~:} Example of an attractor cycle of the path of characteristics for $m=2$, $\omega_p=-3.75$, solid-body rotation ($\varepsilon=0$) and the solar aspect ratio $\eta=0.71$. }
\label{fig:attractors1}
\end{figure*}

In the case of solid-body rotation, the paths of characteristics in a meridional plane follow straight lines, which form a constant angle with the rotation axis --- \emph{i.e.} the $z$-axis ---, that only depends on the (uniform) Doppler-shifted wavefrequency $\tilde{\omega}_p = \omega_p+m$ (see the dispersion relation given in Eq. \ref{eq:dispersion2} for $\varepsilon = 0$). For a given aspect ratio of the shell $\eta$, certain values of the frequency $\omega_p$ cause the paths of characteristics to converge towards periodic orbits called ``attractors'', while for some other frequencies, paths of characteristics nearly occupy the whole shell with no obvious periodic pattern. These two different behaviors are illustrated in Fig. \ref{fig:attractors1}. The existence of one or several attractor(s) for a given set of parameters may be investigated by studying the exponential convergence rate of two close characteristics along multiple reflexions. This is quantified by the Lyapunov exponent $\Lambda$ \citep[for more details, see][]{Rieutord2001, BR2013}. The Lyapunov exponent can be defined as
\begin{equation}
\Lambda = \lim_{N \rightarrow \infty} \frac{1}{N} \sum_{k=1}^N \ln \left|\frac{\mathrm{d}x_{k+1}}{\mathrm{d}x_k}\right|,
\label{eq:lyapunov}
\end{equation}
where $\mathrm{d}x_k$ is the separation between two characteristics after $k$ reflections. In our case, this quantity can be evaluated using either the reflexions on the equator ($\mathrm{d}s_k$), on the rotation axis ($\mathrm{d}z_k$) or on the inner (resp. outer) boundary of the shell $\mathrm{d}\theta^{in}_k$ (resp. $\mathrm{d}\theta^{out}_k$). $\Lambda \approx 0$ corresponds to the case of ``space-filling'' paths of characteristics, whereas $\Lambda<0$ identifies the existence of an attractor (see left and right panel of Fig. \ref{fig:attractors1} respectively).

In the case of solid-body rotation, the locations where characteristics bounce off the shell boundaries, the equator and the rotation axis can be determined analytically, thus allowing for a semi-analytic calculation of $\Lambda$ \citep{Rieutord2001}. However, differential rotation requires the numerical integration of the path of characteristics. For that purpose, we choose a starting point in the shell and we follow the propagation by numerically integrating Eq. (\ref{eq:slope2}). Since characteristics cannot propagate in the elliptic domain, it is necessary to impose reflexion at turning surfaces.

\begin{figure}
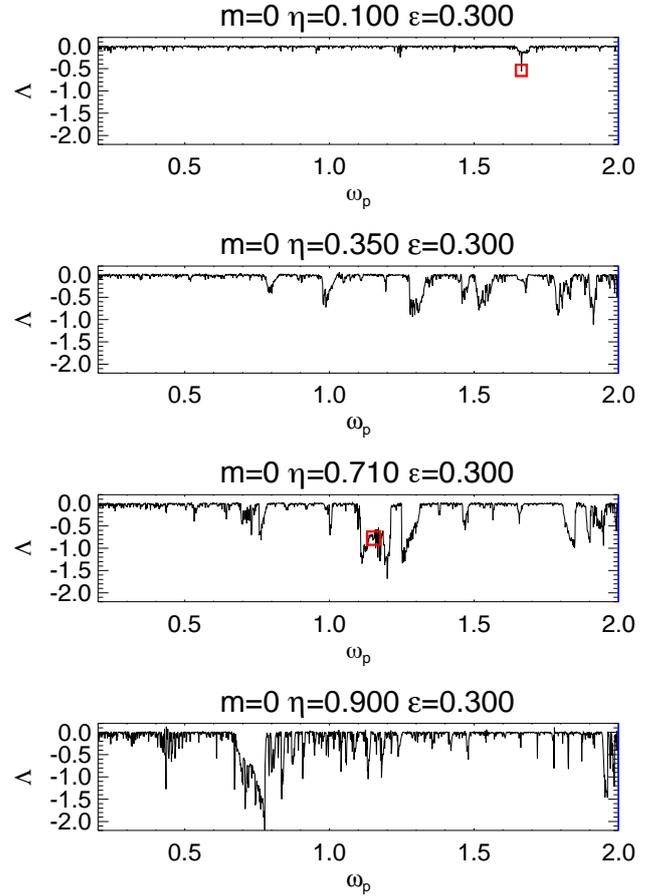

\centering
\includegraphics[width=0.49\textwidth]{{{lyapunovmin_m0_eps0.30_eta0.10}}}
\includegraphics[width=0.49\textwidth]{{{lyapunovmin_m0_eps0.30_eta0.35}}}
\includegraphics[width=0.49\textwidth]{{{lyapunovmin_m0_eps0.30_eta0.71}}}
\includegraphics[width=0.49\textwidth]{{{lyapunovmin_m0_eps0.30_eta0.90}}}
\caption{Numerical spectrum of the Lyapunov exponent for $m=0$, $\varepsilon=0.30$ and $\omega_p \in [0.2, 2.0]$. From the top to bottom panels, $\eta=\{0.10, 0.35, 0.71, 0.90\}$. The red open squares correspond to the modes shown in Figs. \ref{fig:Dmode_2} to \ref{fig:Dmode_nocore_noattractor}.}
\label{fig:lyapunov_eta}
\end{figure}

\begin{figure}
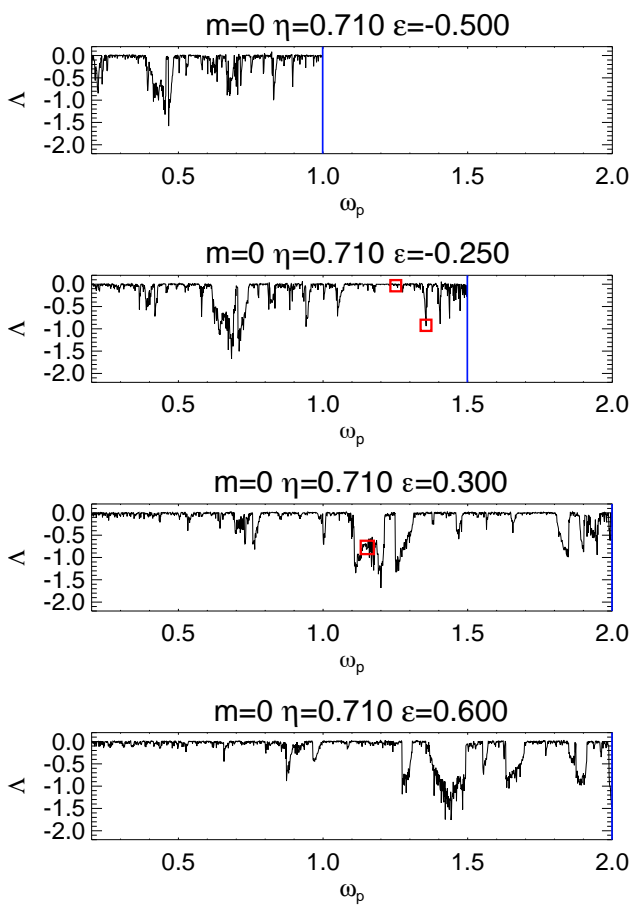

\centering
\includegraphics[width=0.49\textwidth]{{{lyapunovmin_m0_eps-0.50_eta0.71}}}
\includegraphics[width=0.49\textwidth]{{{lyapunovmin_m0_eps-0.25_eta0.71}}}
\includegraphics[width=0.49\textwidth]{{{lyapunovmin_m0_eps0.30_eta0.71}}}
\includegraphics[width=0.49\textwidth]{{{lyapunovmin_m0_eps0.60_eta0.71}}}
\caption{Numerical spectrum of the Lyapunov exponent for $m=0$, $\eta=0.71$ and $\omega_p \in [0.2, 2.0]$. From the top to bottom panels, $\varepsilon=\{-0.50, -0.25, 0.30, 0.60\}$. The blue solid line marks the frequency of the transition between D and DT modes. The red open squares correspond to the modes shown in Figs. \ref{fig:Dmode_2} to \ref{fig:Dmode_nocore_noattractor}.}
\label{fig:lyapunov_eps}
\end{figure}

For illustration purposes, we evaluated numerically the Lyapunov exponents as a function of the eigenfrequency for $m=0$ for different values of $\eta$ and $\varepsilon$, more precisely :
\begin{itemize}
\item $\eta=\{0.10, 0.35, 0.71, 0.90\}$ along with a fixed solar-like value of the differential rotation parameter $\varepsilon=0.30$ for Fig. \ref{fig:lyapunov_eta},
\item $\varepsilon=\{-0.50, -0.25, 0.30, 0.60\}$ along with a fixed solar-like value of the convective shell aspect ratio $\eta=0.71$ for Fig. \ref{fig:lyapunov_eps}.
\end{itemize}

We used 800 frequencies per unit interval of frequencies and ten pairs of characteristics for each data point before averaging the results. The results shown in Figs. \ref{fig:lyapunov_eta} and \ref{fig:lyapunov_eps} should therefore be considered rather as a qualitative indicator of the presence of short-period attractor cycles than a quantitative one because our method probably yields substantial uncertainties on the values of $\Lambda$. Note also that the focusing of characteristics towards a wedge in the frequency range of DT modes (see the bottom-left panel of Fig. \ref{fig:attractors2} below) prevented us from computing values of $\Lambda$ for the DT modes. As expected, attractors of various strengths exist in small intervals of frequencies where $\Lambda$ is strictly negative whereas intervals where $\Lambda \approx 0$ are associated with very long-period attractors or quasi-periodic orbits of characteristics.

\begin{figure*}
\centering
\includegraphics[width=0.4\textwidth]{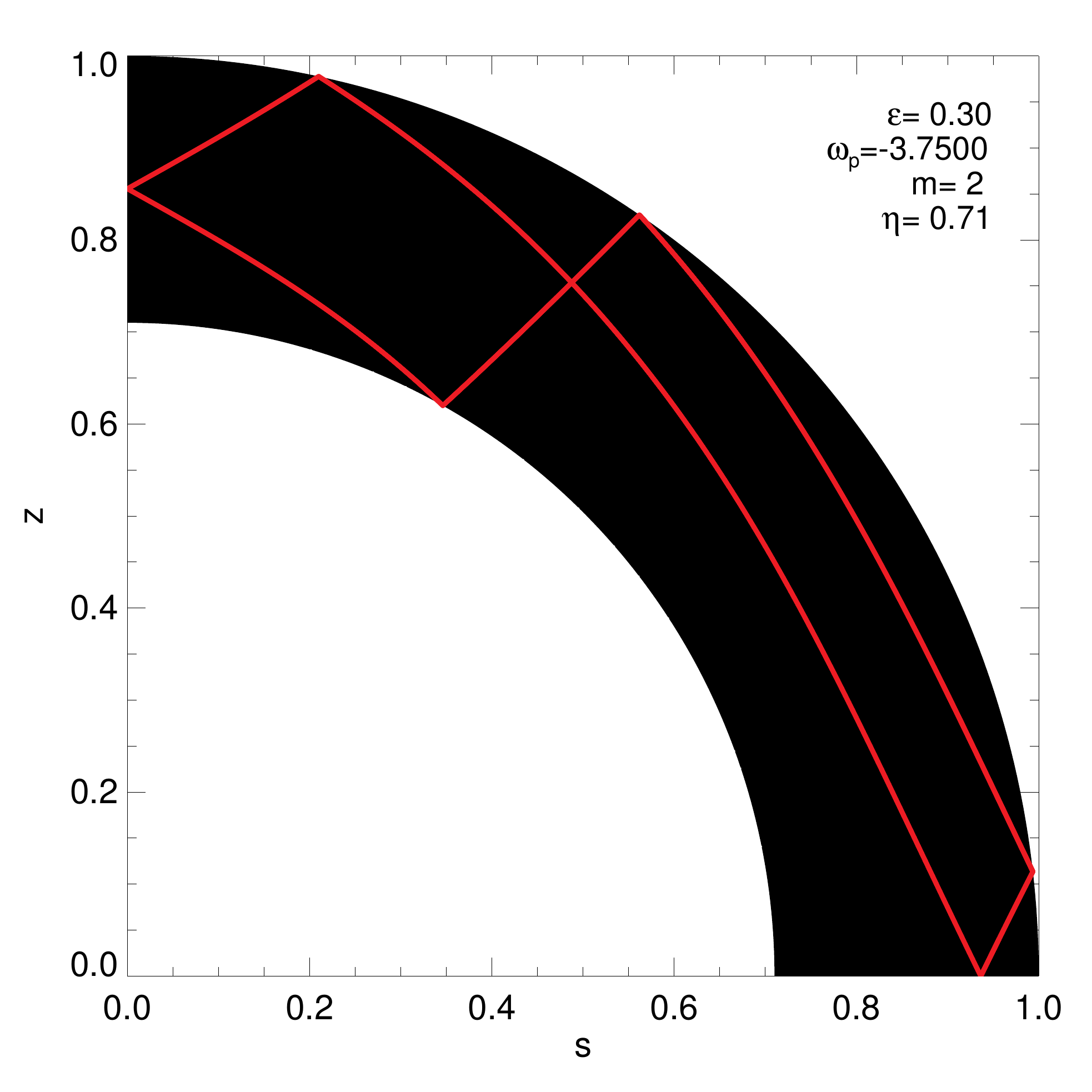} \qquad
\includegraphics[width=0.4\textwidth]{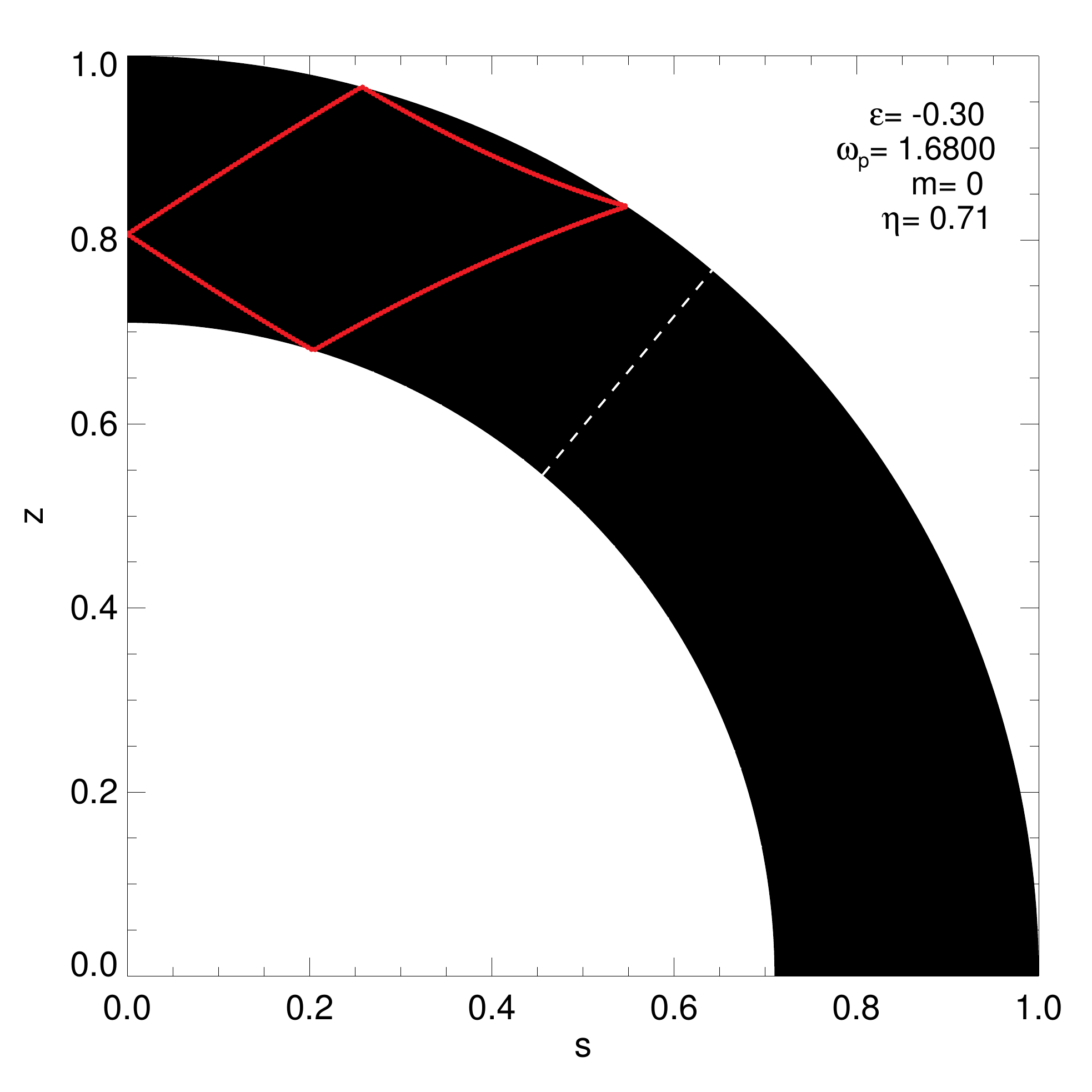}
\includegraphics[width=0.4\textwidth]{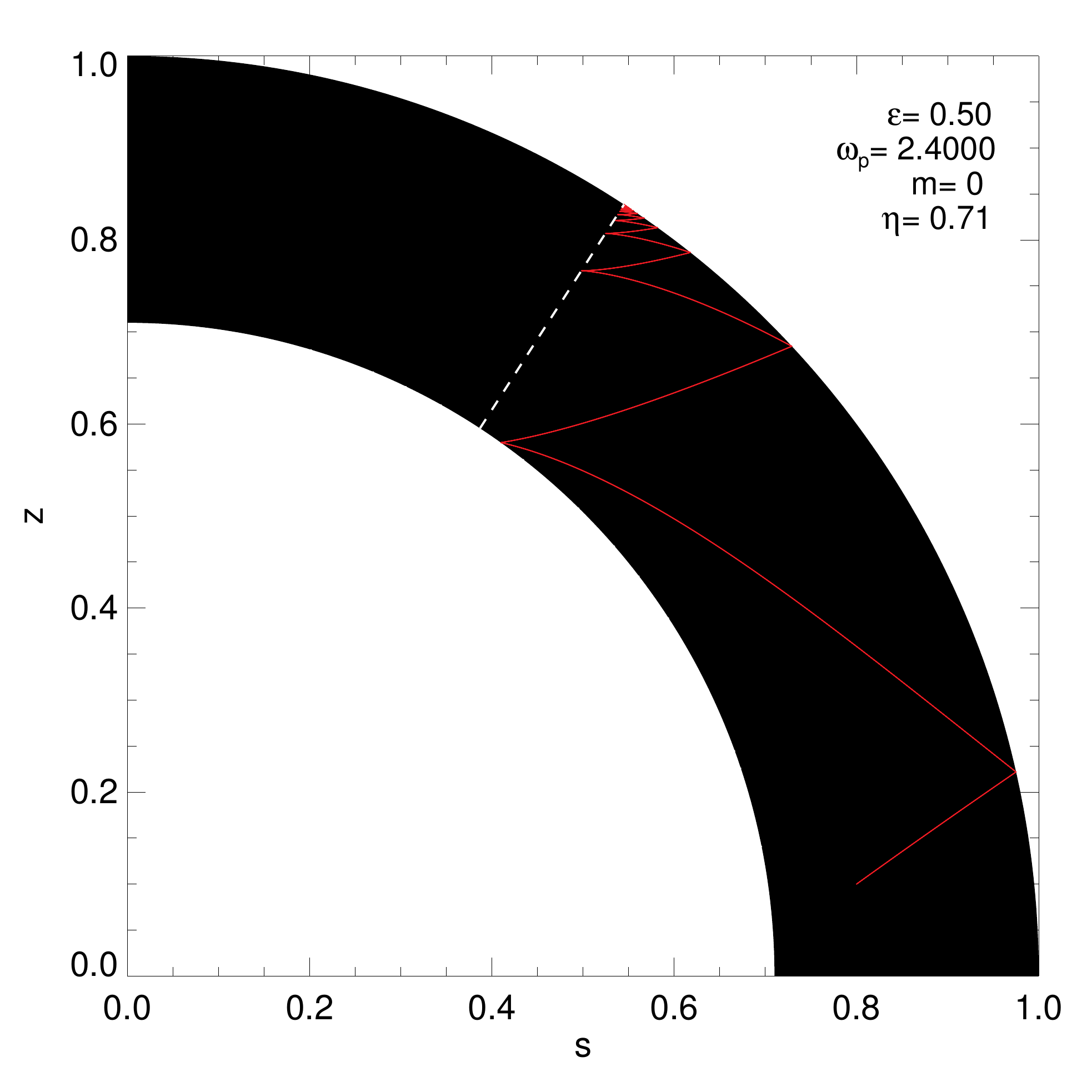} \qquad
\includegraphics[width=0.4\textwidth]{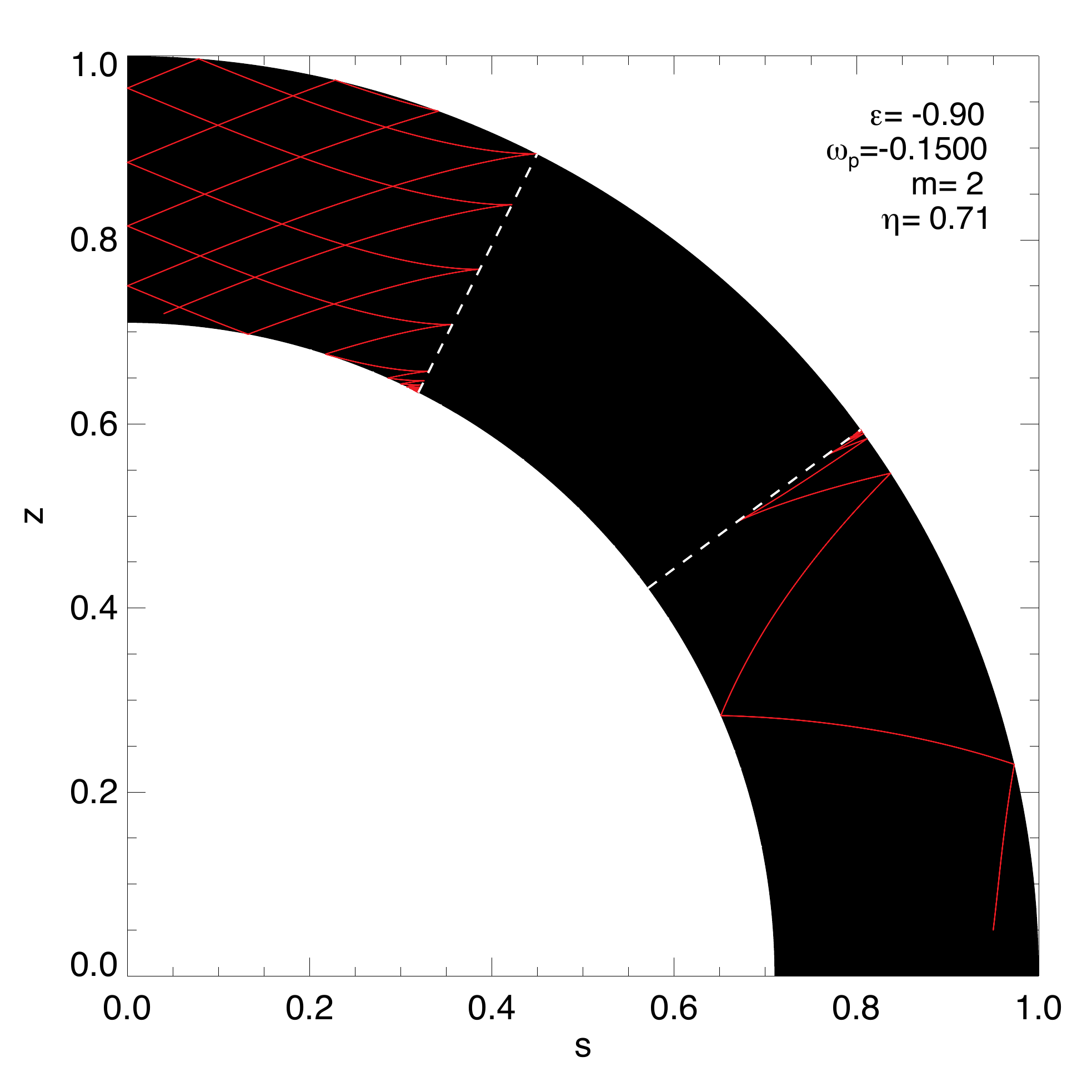}
\caption{{\bf Upper left~:} Example of attractor cycle for a D mode with $m=2$, $\omega_p=-3.75$ in a solar-like convective envelope ($\varepsilon=0.3$, $ \eta=0.71$). {\bf Upper right~:} Same for a DT mode with $m=0$, frequency $\omega_p=1.68$ and anti-solar conical rotation ($\varepsilon = -0.3$), the white dashed line shows the turning surface. {\bf Bottom left~:} Illustration of the focusing of the paths of characteristics at the intersection of the turning surface (dashed line) and the outer boundary of the shell ($m=0$, $\omega_p=2.4$ and $\varepsilon=0.5$). {\bf Bottom right~:} For $m=2$, $\omega_p=-0.126$ and $\varepsilon=-0.90$, two turning surfaces exist within the shell, allowing the propagation of characteristics near the equator or the poles but not in between.}
\label{fig:attractors2}
\end{figure*}

Fig. \ref{fig:lyapunov_eta} illustrates how the existence of short-period attractors in the frequency range of inertial waves can be affected by the size of the inner core. It seems that when the core is small, attractors of characteristics are very rare but still exist as illustrated by the region around $\omega_p \approx 1.66$ in the top-left panel of Fig. \ref{fig:lyapunov_eta} . We checked that this attractor exists regardless of the size of the inner core as long as the latter is small enough, and we computed the corresponding eigenmode (see Fig. \ref{fig:Dmode_nocore_noattractor}, and Fig. 8 in \cite{BR2013}). However, the other panels of Fig. \ref{fig:lyapunov_eta} indicate that more and more attractors exist with increasing the value of $\eta$, due to increasingly numerous reflections on the inner core. The case where $\eta = 0.90$ shows that a large number of short period attractors exist over the frequency range of D modes.

Fig. \ref{fig:lyapunov_eps} shows the evolution of the Lyapunov exponent spectrum for a given core size but for various values of the differential rotation parameter $\varepsilon$. We see that the occurrence of short-period attractors is not sensitive to $\varepsilon$ but some features of the Lyapunov spectra can be tracked as they are progressively altered and shifted to higher frequencies with increasing $\varepsilon$ --- for instance, the deep double-peak starting from $\omega_p \in [0.3, 0.5]$ in the top-left panel and shifted to $\omega_p \in [1.25, 1.5]$ in the bottom-right panel.

When the fluid is differentially rotating with angular velocity $\Omega(\theta)$, paths of characteristics may still converge towards attractors --- just like in the case of solid-body rotation --- but with curved characteristics as illustrated in the top-left panel of Fig. \ref{fig:attractors2}. They also depend on the azimuthal wavenumber $m$ through the non-uniform Doppler-shifted frequency $\tilde{\omega}_p$. We can also find attractors of characteristics when a turning surface exists in the shell (top-right panel). Sometimes, characteristics tend to focus at the intersection of a turning surface with the boundaries of the shell as illustrated by the bottom-left plot of Fig. \ref{fig:attractors2}, showing a behavior that is similar to the one found by \cite{Dintrans1999} for gravito-inertial waves. Finally, we emphasize that in rare cases, two turning surfaces can exist within the shell, affecting the waves' propagation domain accordingly : it is either restricted to the polar and equatorial regions (as illustrated in the bottom-right panel of Fig. \ref{fig:attractors2}) or restricted to mid-latitudinal regions.

\section{Viscous problem: shear layers, comparison to inviscid analysis, behaviour at corotation resonances}
\label{sec:viscous_problem}

Our study of the propagation properties of inertial waves in a differentially rotating inviscid fluid with a conical rotation profile (see section \ref{sec:inviscid_problem}) shows that two families of inertial modes of oscillation exist : D modes that may propagate in the entire shell, and DT modes that are trapped in part of the shell because of the existence of a turning surface. We find that paths of characteristics depend on $m$ through the Doppler-shifted wavefrequency, which is not the case for solid-body rotation, but the main difference between $m=0$ and $m\neq0$ modes is the possible existence of corotation resonances in the latter case.

In Sect. \ref{sec:numericalmethod}, we briefly present the numerical method we used in order to solve the viscous problem exposed in Sect. \ref{sec:physical_model}. Then we study in Sect. \ref{sec:D} (respectively Sect. \ref{sec:DT}) the general properties of weakly-damped viscous D modes that propagate in the whole spherical shell (resp. DT modes that propagate in part of the shell) via numerical simulations. We show in particular how the shear layer structure can be related to the inviscid analysis detailed above in Sect. \ref{sec:inviscid_problem}. As explained in the introduction of this work, we will mostly focus on singular modes for which a short-period attractor of characteristics exists since they may be related with strong viscous dissipation \citep{Ogilvie2005}. Finally, we examine $m\neq0$ modes with corotation resonances in Sect. \ref{sec:corotation}.

\subsection{Numerical method}
\label{sec:numericalmethod}

The linearized dimensionless system of Eqs. (\ref{eq:viscous_problem}) and the associated stress-free boundary conditions are solved using a unique decomposition of the unknown velocity field $\bf u$ onto vectorial spherical harmonics \citep{Rieutord1987} :
\begin{multline}
\label{eq:decomposition}
{\bf u}(x,\theta,\varphi) = \sum_{l=0}^{\infty} \sum_{m=-l}^{l} \left\{ u^l_m(x) \, {\bf R}^m_l(\theta,\varphi) + v^l_m(x) \, {\bf S}^m_l(\theta,\varphi) \right. \\
\left. + w^l_m(x) \, {\bf T}^m_l(\theta,\varphi) \right\}
\end{multline}
with ${\bf R}^m_l = Y_l^m(\theta,\varphi) \, {\bf e_r}$, ${\bf S}^m_l = {\bf \nabla_H} Y_l^m$ and ${\bf T}^m_l = {\bf \nabla_H} \times {\bf R}^m_l$ where $Y^m_l(\theta, \varphi)$ is the usual spherical harmonic of degree $l$ and order $m$ normalized on the unit sphere, and ${\bf \nabla_H} = {\bf e_{\theta}} \, \partial_{\theta} + {\bf e_{\varphi}} (\sin\theta)^{-1} \, \partial_{\varphi} $ is the horizontal gradient.

The projection of the continuity equation on $Y^m_l$ gives $v^l_m$ as a simple function of $u^l_m$ and $du^l_m/dx$. The momentum equation is then projected on ${\bf R}^m_l$ and ${\bf T}^m_l$, which yields two long ordinary differential equations involving only the radial functions $u^l_m$ and $w^l_m$ (since $v^l_m$ can be eliminated) :
\begin{enumerate}
\item the right hand side of the second-order equation satisfied by $w^l_m$ is a linear combination of $u^{l\pm3}_m$, $du^{l\pm3}_m/dx$, $u^{l\pm1}_m$, $du^{l\pm1}_m/dx$ and $w^{l\pm2}_m$ (this term vanishes for $m=0$) ;
\item the right-hand side of the fourth-order differential equation satisfied by $u^l_m$ is a linear combination of $w^{l\pm3}_m$, $dw^{l\pm3}_m/dx$, $w^{l\pm1}_m$, $dw^{l\pm1}_m/dx$, $u^{l\pm2}_m$, $du^{l\pm2}_m/dx$ and $d^2u^{l\pm2}_m/dx^2$ (these last three terms vanish for $m=0$). 
\end{enumerate}

The fact that the radial functions $u^l_m$, $v^l_m$ and $w^l_m$ with different azimuthal wavenumbers $m$ are not coupled is a consequence of the axisymmetry of the background flow. The coupling between different latitudinal degrees (from $l-3$ to $l+3$) results from the choice of our specific conical rotation profile through the Coriolis acceleration terms.

In order to solve these equations numerically, they are discretized in the radial direction on the Gauss-Lobatto collocation nodes associated with Chebyshev polynomials. Each set of equations is truncated at order $L$ for the spherical harmonics basis, and at order $N_r$ for the Chebyshev basis. This yields a finite eigenvalue problem of order $N_r \times L$ with a block banded matrix composed of (up to) seven block bands. Then we use the linear solver based on the incomplete Arnoldi-Chebyshev algorithm \citep[see details in][]{Valdettaro2007} to compute pairs of eigenvalues (equal to $i \omega_p$) and eigenvectors (values of $u^l_m$, $v^l_m$ and $w^l_m$ on each point of the radial grid), given an initial value guess for $i \omega_p$. Since the values of $N_r$ and $L$ required to achieve spectral convergence vary a lot from one mode to another, especially for the very demanding small values of $E$, some figures will be accompanied by the spectral content of the velocity field. In all the cases presented in this work we assumed symmetry with respect to the equatorial plane which explains that our results are shown only for positive values of $z$, but anti-symmetry is also possible.

\subsection{Axisymmetric and non-axisymmetric D modes with no corotation layer}
\label{sec:D}

\begin{figure*}
\centering
\includegraphics[width=0.4\textwidth]{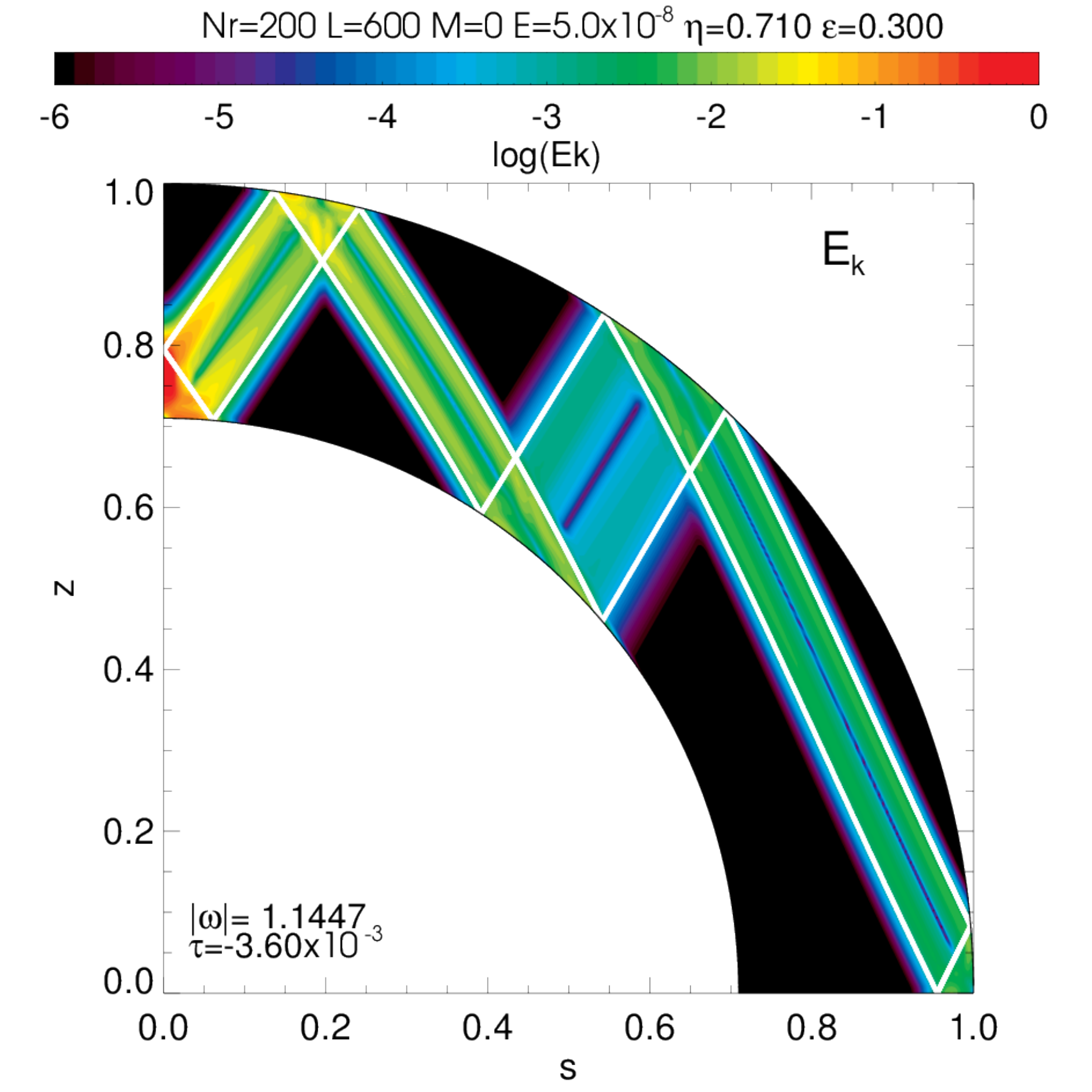}
\includegraphics[width=0.5\textwidth]{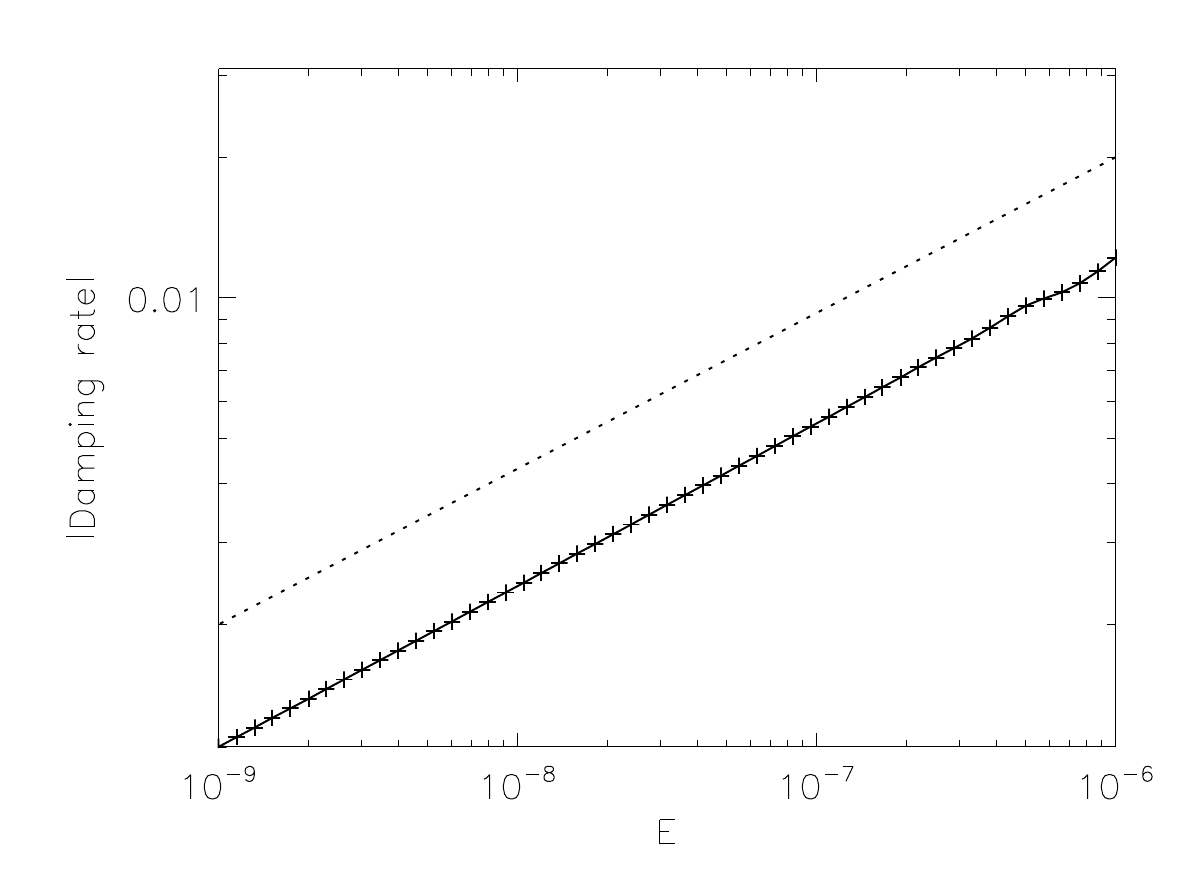}
\caption{{\bf Left~:} Meridional cut of the normalised kinetic energy of an axisymmetric D mode with eigenfrequency $\omega_p\approx1.15$ obtained with $E=5\times10^{-8}$, the Sun's aspect ratio $\eta=0.71$ and conical differential rotation $\varepsilon=0.3$. The attractor of characteristics for these parameters is overplotted by the white curve. {\bf Right~:} Scaling of the damping rate with respect to the Ekman number $E$. The dashed line is proportional to $E^{1/3}$.}
\label{fig:Dmode_2}
\end{figure*}

\begin{figure*}
\centering
\includegraphics[width=0.4\textwidth]{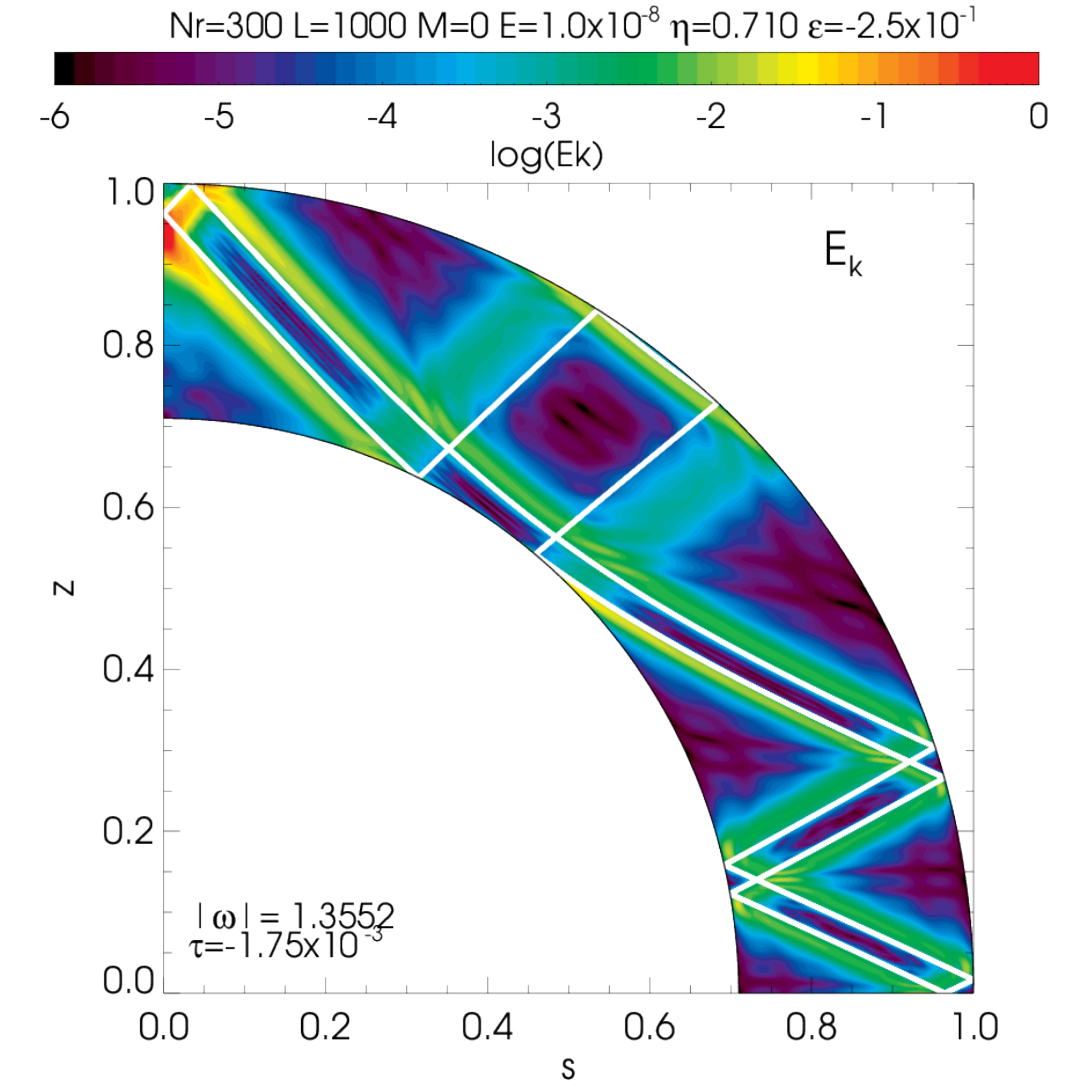} \qquad
\includegraphics[width=0.4\textwidth]{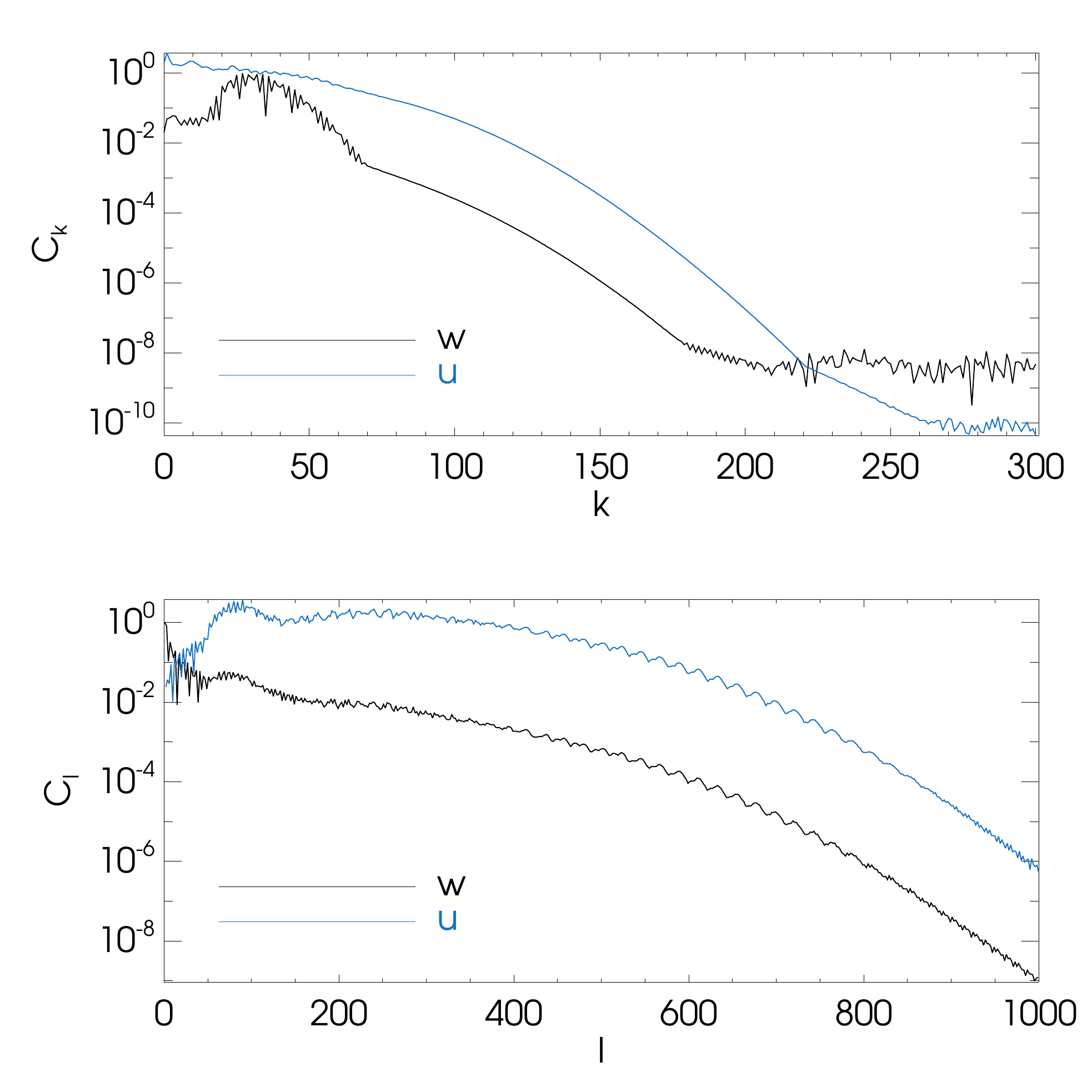}
\caption{{\bf Left~:} Meridional cut of the normalised kinetic energy of a D mode with eigenfrequency $\omega_p\approx1.355$ obtained with $E=10^{-8}$ and anti-solar differential rotation $\varepsilon=-0.25$. The attractor of characteristics is overplotted by the thick white curve. {\bf Right~:} Spectral content of the radial ($u$) and orthoradial ($w$) components of the velocity field for this mode. Chebyshev and spherical harmonics coefficients are shown in the top and bottom panels respectively.}
\label{fig:Dmode_1}
\end{figure*}

\begin{figure*}
\centering
\includegraphics[width=0.4\textwidth]{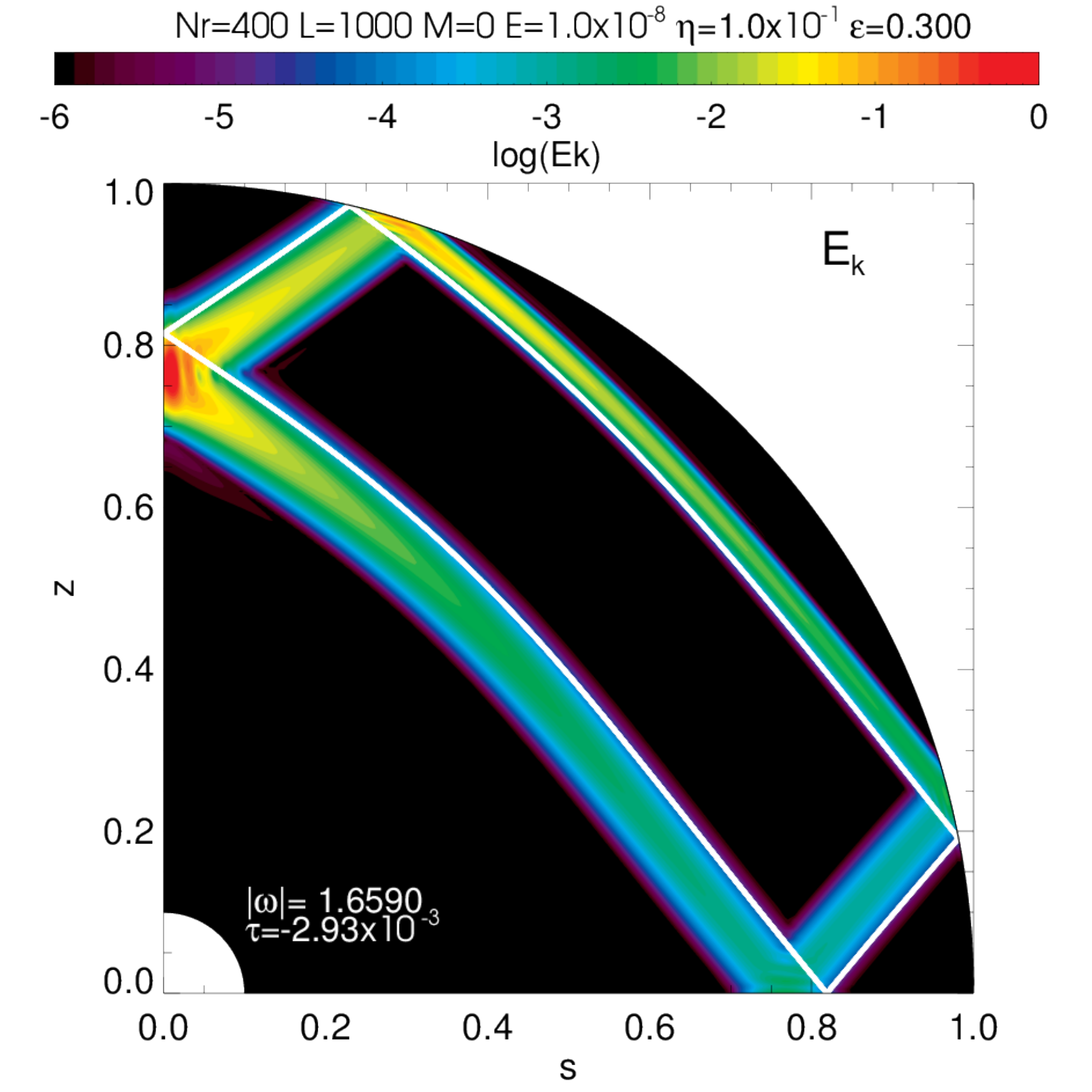} \qquad
\includegraphics[width=0.4\textwidth]{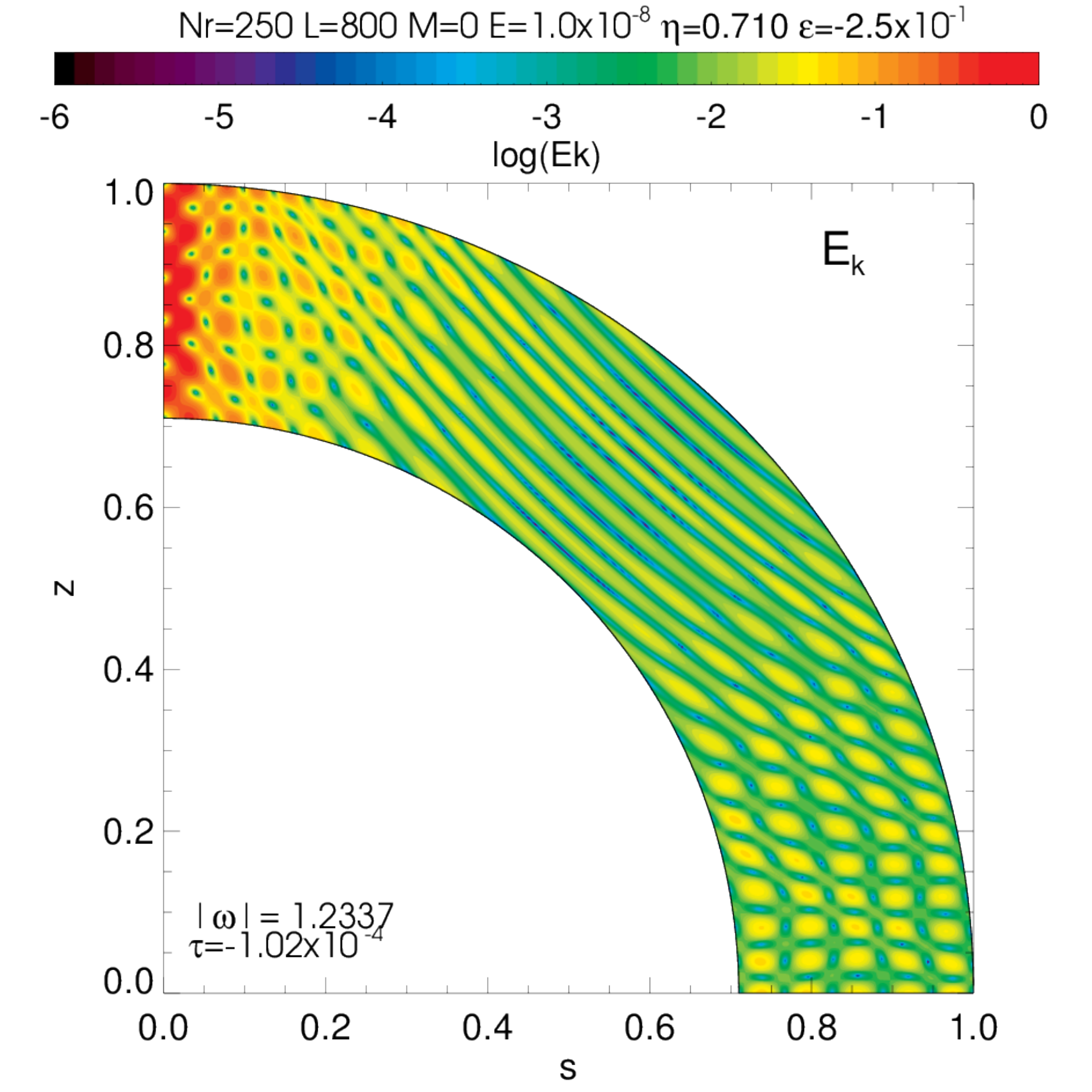}
\caption{{\bf Left~:} Meridional cut of the normalised kinetic energy of a D mode with eigenfrequency $\omega_p\approx1.66$ obtained with $E=10^{-8}$, $\eta=0.10$ and conical differential rotation $\varepsilon=0.30$. The attractor of characteristics is again overplotted by the thick white curve. {\bf Right~:} Meridional cut of the normalised kinetic energy of a D mode with eigenfrequency $\omega_p\approx1.23$ obtained with $E=10^{-8}$, $\eta=0.71$ and conical differential rotation $\varepsilon=-0.25$. This mode is clearly associated with a quasi-periodic orbit of characteristics and not with an attractor.}
\label{fig:Dmode_nocore_noattractor}
\end{figure*}

We present in this section our numerical results for a few representative D modes with a background conical rotation profile and Ekman numbers between $10^{-7}$ and $10^{-8}$, and we compare the structural properties of the shear layers with the inviscid analysis detailed in Sect. \ref{sec:inviscid_problem}. Note that in the rest of this paper, we always show the kinetic energy distribution of the computed velocity fields, but that the viscous dissipation looks qualitatively very similar \citep[e.g. Fig. 13 in][]{BR2013}.

Fig. \ref{fig:Dmode_2} displays the result of one of our numerical calculations for an axisymmetric ($m=0$) D mode with eigenfrequency $\omega_p\approx 0.91$ obtained in a spherical shell of solar aspect ratio $\eta = 0.71$ with solar-like conical differential rotation ($\varepsilon=0.30$). The left panel displays the spatial distribution of the mode's kinetic energy in a meridional quarter-plane. The amplitude of the mode is maximum near the rotation axis which is a feature shared by inertial modes in a uniformly rotating fluid shell. This has been demonstrated in detail in the appendix A of \cite{Rieutord1997} who showed that the kinetic energy along characteristic trajectories grows as $s^{-1/2}$ when $s \rightarrow 0$. The structure of this specific mode mostly consists of a shear layer following a short-period attractor formed by (slightly) curved lines, as expected in the case of differential rotation (see Sect. \ref{sec:inviscid_problem}). This attractor is overplotted with a thick white curve, which is the prediction for the propagation of characteristics under the short-wavelength approximation. The patterns formed by the attractor and by the shear layers of the viscous mode are in very good agreement. The mode shown in Fig. \ref{fig:Dmode_2} was extracted from a sequence of calculations in which we followed a particular mode while progressively decreasing the Ekman number from $10^{-6}$ to $10^{-9}$. The damping rate of this mode is displayed as a function of $E$ in the right-panel of Fig. \ref{fig:Dmode_2} where the dashed line clearly shows that it is proportional to $E^{1/3}$, which is the scaling that is expected in the asymptotic regime for solid-body rotation \citep{Rieutord1997}, meaning that this kind of D mode is not deeply affected by differential rotation.

We show in Fig. \ref{fig:Dmode_1} a qualitatively similar axisymmetric D mode of eigenfrequency $\omega_p\approx 1.35$ that was obtained with the same aspect ratio but slightly lower Ekman number $E=10^{-8}$ and anti-solar differential rotation $\varepsilon=-0.25$. This time the amplitude of the mode is still maximum at the rotation axis but is also quite large near the inner critical latitude (where the characteristics are tangent to the inner core), which is reminiscent of the solid-body rotation case where shear layers are sometimes emitted at the critical latitude. The right panel of Fig. \ref{fig:Dmode_1} depicts the spectral content of $u$ and $w$ for this mode : the top panel shows the maximum Chebyshev coefficients $C_k$ as a function of the Chebyshev order $k$, taking the maximum value among all the spherical harmonics coefficients for a given $k$ ; similarly, the bottom panel shows the maximum spherical harmonics coefficients $C_l$ as a function of the spherical harmonic degree : for a given $l$, we take the maximum value among all the Chebyshev coefficients. Therefore we are certain that numerical convergence is achieved for this mode.

Finally, the axisymmetric mode of frequency $\omega_p\approx1.66$ presented in the left panel of Fig. \ref{fig:Dmode_nocore_noattractor} displays a shear layer that follows an attractor of characteristics that exists for arbitrarily small values of the aspect ratio $\eta$. This means that in conical differential rotation, attractors of characteristics may exist independently of the existence of an inner core, which was also found for cylindrical and shellular differential rotation profiles by \cite{BR2013}. We checked that this mode does exist for arbitrarily small cores, which contrasts with the fact that inertial modes in a full sphere are regular in the case of solid-body rotation \citep{Greenspan1968, Zhang2001}. Our result shows that this is probably no longer the case when differential rotation is taken into account. The case where characteristics do not converge towards any limit cycle is shown in the right-panel of Fig. \ref{fig:Dmode_nocore_noattractor}, which displays a mode of frequency $\omega_p\approx1.23$ with $\eta=0.71$ and $\varepsilon=-0.25$. This set of parameters correspond to the open square in Fig. \ref{fig:lyapunov_eps} for which $\Lambda \approx 0$. As expected, the shear layer patterns visible here follow the propagation of characteristics so that the kinetic energy is more smoothly distributed than in the previous attractor cases.

\begin{figure*}
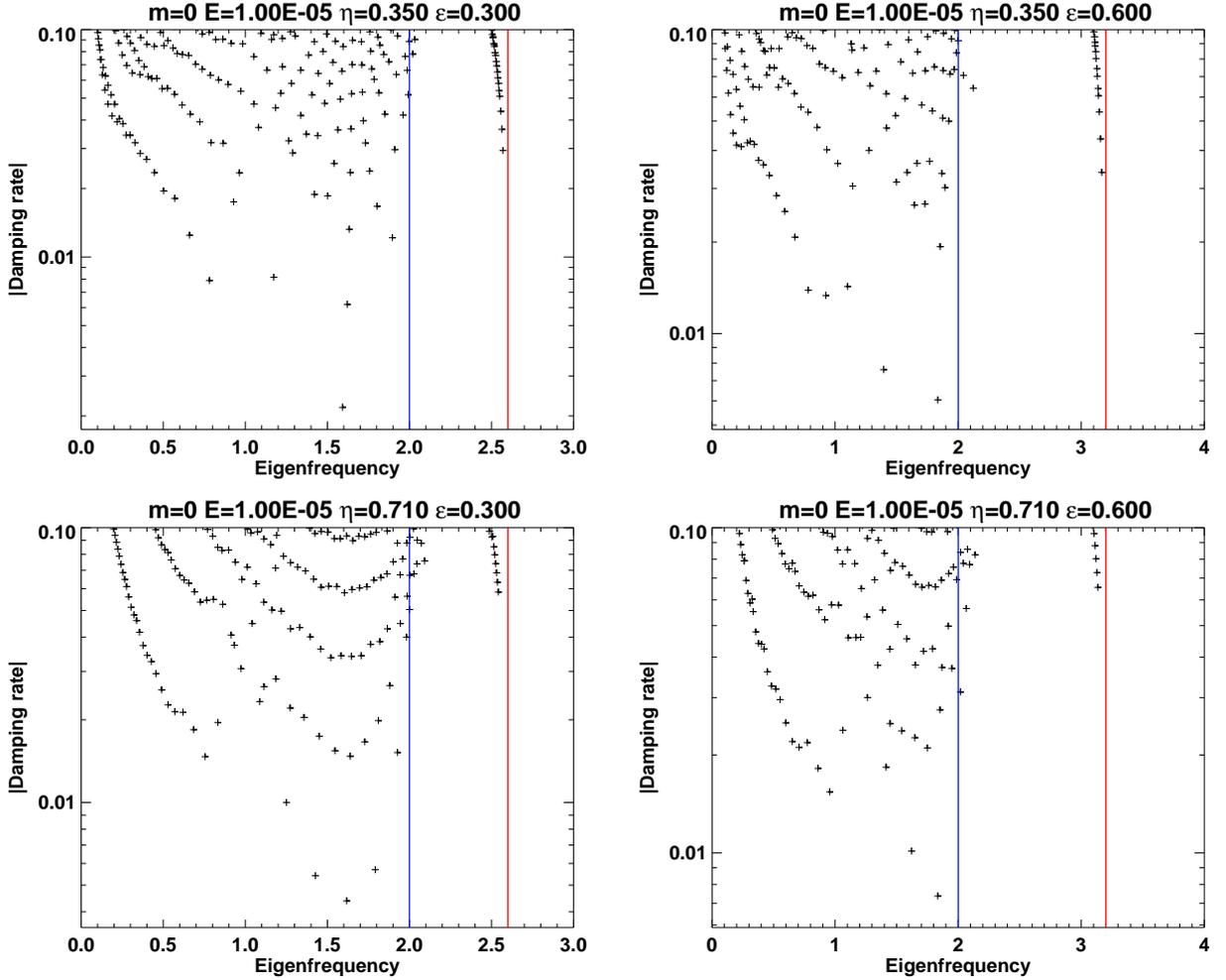

\centering
\includegraphics[width=0.45\textwidth]{{{QZ_m0_eps0.300_eta0.350_e1.00D-05}}}
\includegraphics[width=0.45\textwidth]{{{QZ_m0_eps0.600_eta0.350_e1.00D-05}}}
\includegraphics[width=0.45\textwidth]{{{QZ_m0_eps0.300_eta0.710_e1.00D-05}}}
\includegraphics[width=0.45\textwidth]{{{QZ_m0_eps0.600_eta0.710_e1.00D-05}}}
\caption{Distribution of eigenvalues in the complex plane for $m=0$ and $E=10^{-5}$, with a spectral resolution of $N_r=60$ and $L=150$. The top and bottom rows are for thick and thin shells ($\eta=0.35$ and $\eta=0.71$ respectively) while the left and right columns are for $\varepsilon=0.30$ and $\varepsilon=0.60$. In all panels, the vertical blue (resp. red) line depicts the transition between the D and DT modes (resp. DT and non-existant modes) frequency ranges.}
\label{fig:QZ}
\end{figure*}

The four modes shown in Figs. \ref{fig:Dmode_2} to \ref{fig:Dmode_nocore_noattractor} all correspond to one of the open squares in Figs. \ref{fig:lyapunov_eta} and \ref{fig:lyapunov_eps}. This highlights the interest of the inviscid approach undertaken in Sect. \ref{sec:inviscid_problem} since the kinetic energy of a given mode is indeed more localized along thin shear layers when $\Lambda < 0$ and more uniformly-distributed in the shell when $\Lambda \approx 0$. As a conclusion, the inviscid analysis allows us to understand how differential rotation affects inertial modes, although we restricted ourselves to D modes that can propagate in the whole spherical shell. We see that D modes behave quite similarly to inertial modes for solid-body rotation. In the next section we turn to DT modes, which are specific to differential rotation.

\begin{figure*}
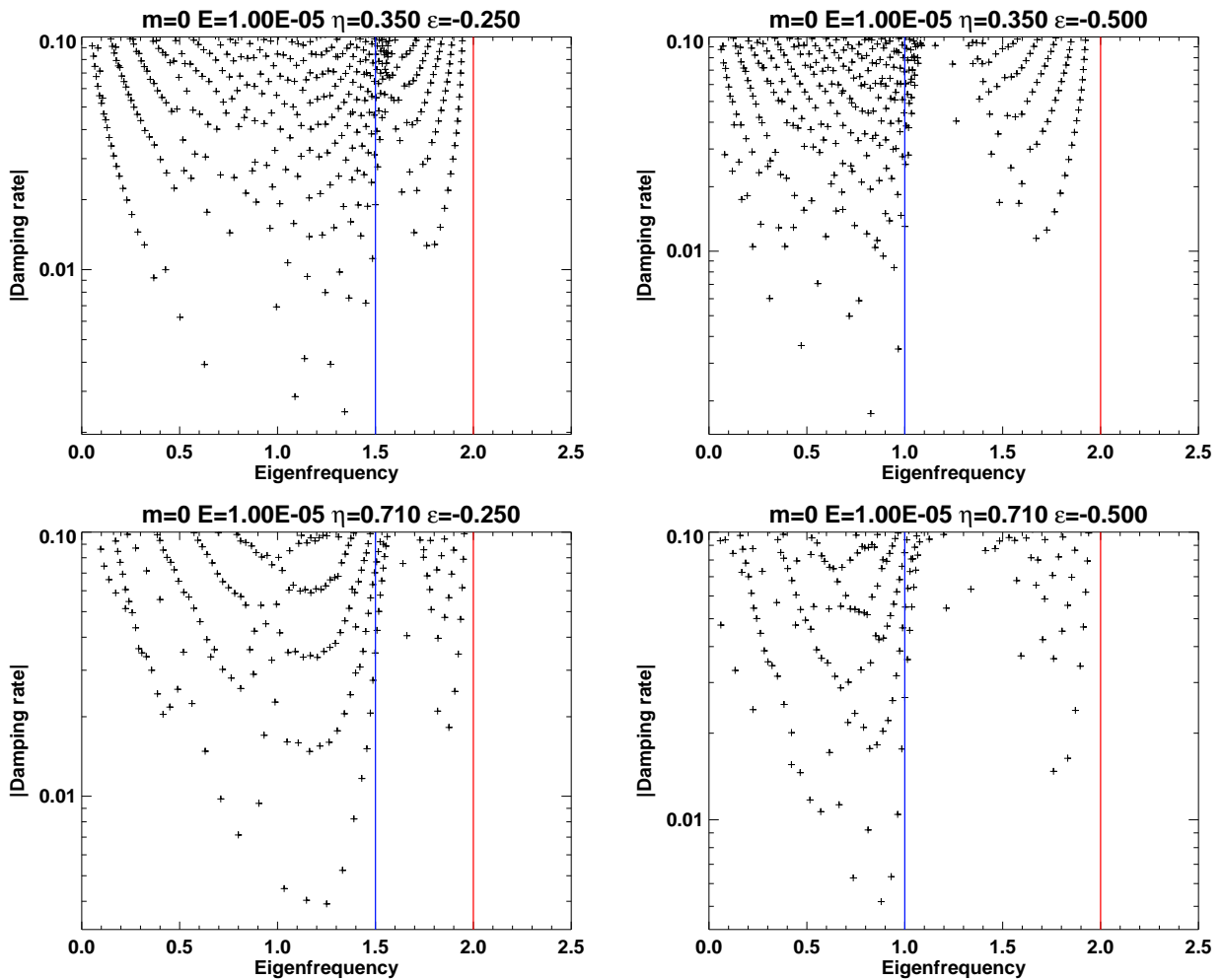

\centering
\includegraphics[width=0.45\textwidth]{{{QZ_m0_eps-0.250_eta0.350_e1.00D-05}}}
\includegraphics[width=0.45\textwidth]{{{QZ_m0_eps-0.500_eta0.350_e1.00D-05}}}
\includegraphics[width=0.45\textwidth]{{{QZ_m0_eps-0.250_eta0.710_e1.00D-05}}}
\includegraphics[width=0.45\textwidth]{{{QZ_m0_eps-0.500_eta0.710_e1.00D-05}}}
\caption{Same as Fig. \ref{fig:QZ} except that the left and right columns are for $\varepsilon=-0.25$ and $\varepsilon=-0.50$.}
\label{fig:QZ_2}
\end{figure*}

\subsection{Axisymmetric and non-axisymmetric DT modes}
\label{sec:DT}

In this section, we focus on modes for which a turning surface exists in the shell. The frequency range in which these modes may exist is indicated by the white areas in Fig. \ref{fig:BBR}. However, not all sets of parameters falling in these white areas effectively correspond to an eigenmode. The inviscid analysis exposed in Sect. \ref{sec:inviscid_problem} reveals that characteristics cannot propagate in the elliptic domain of the shell, \emph{i.e.} where $\xi < 0$ or equivalently $\left|\tilde{\omega_p}(\theta)\right| > 2 \omega(\theta)$. As a consequence --- for sufficiently small Ekman numbers --- DT modes are expected to be trapped in the hyperbolic domain out of which characteristics cannot propagate, a property which has always been verified by our numerical calculations. We also present results in cases where characteristics converge towards the intersection of a turning surface with the inner or outer shell boundaries (see Fig. \ref{fig:attractors2}).

\subsubsection{Existence of DT modes}

In order to find DT modes, as a first step we checked the distribution of the eigenvalues of our linear problem for a given set of parameters, using a QZ factorisation method. This method rapidly becomes very costly as the spectral resolution (and thus the dimensions of the linear problem) increases and therefore can only be performed at a low spectral resolution, effectively limiting us to moderate values of $E$. Despite this limitation, trends are still discernable, especially with the sign of the differential rotation parameter $\varepsilon$.

Fig. \ref{fig:QZ} displays our results for a few set of parameters with solar rotation ($\varepsilon > 0$) that were obtained with a spectral resolution of $N_r = 60$ and $L=150$. This resolution is usually sufficient to achieve spectral convergence for all the modes we computed at $E = 10^{-5}$. Note that the eigenvalues with absolute damping rates above $\sim 10^{-1}$ should be ignored because they are dominated by round-off errors and do not represent real resonant modes. We checked that the least-damped eigenvalues located in the bottom part of each panel remained identical when slightly increasing or decreasing either $N_r$ or $L$, while the most-damped eigenvalues changed substantially --- as expected.

In all cases presented here, the least-damped modes are all D modes whereas the potential DT modes have a much higher absolute damping rate and tend to be located near the boundaries of the DT-frequency range displayed by the blue and red lines. This stresses that it is numerically much more difficult to compute DT modes than D modes, as we have indeed experienced. We followed one of these potential DT modes, starting with an eigenfrequency close to the maximum frequency allowed for DT modes (\emph{i.e.} the red line in Fig.~\ref{fig:QZ}) and decreasing the Ekman number from $10^{-5}$ to $10^{-9}$. We found that as $E$ decreases, the eigenfrequency converges towards the aforementioned maximum frequency. This frequency shift affects the location of the turning surface and therefore the size of the hyperbolic domain which simply disappears as $E$ decreases. The very existence of this DT mode at lower Ekman numbers is therefore questionable.

\begin{figure*}
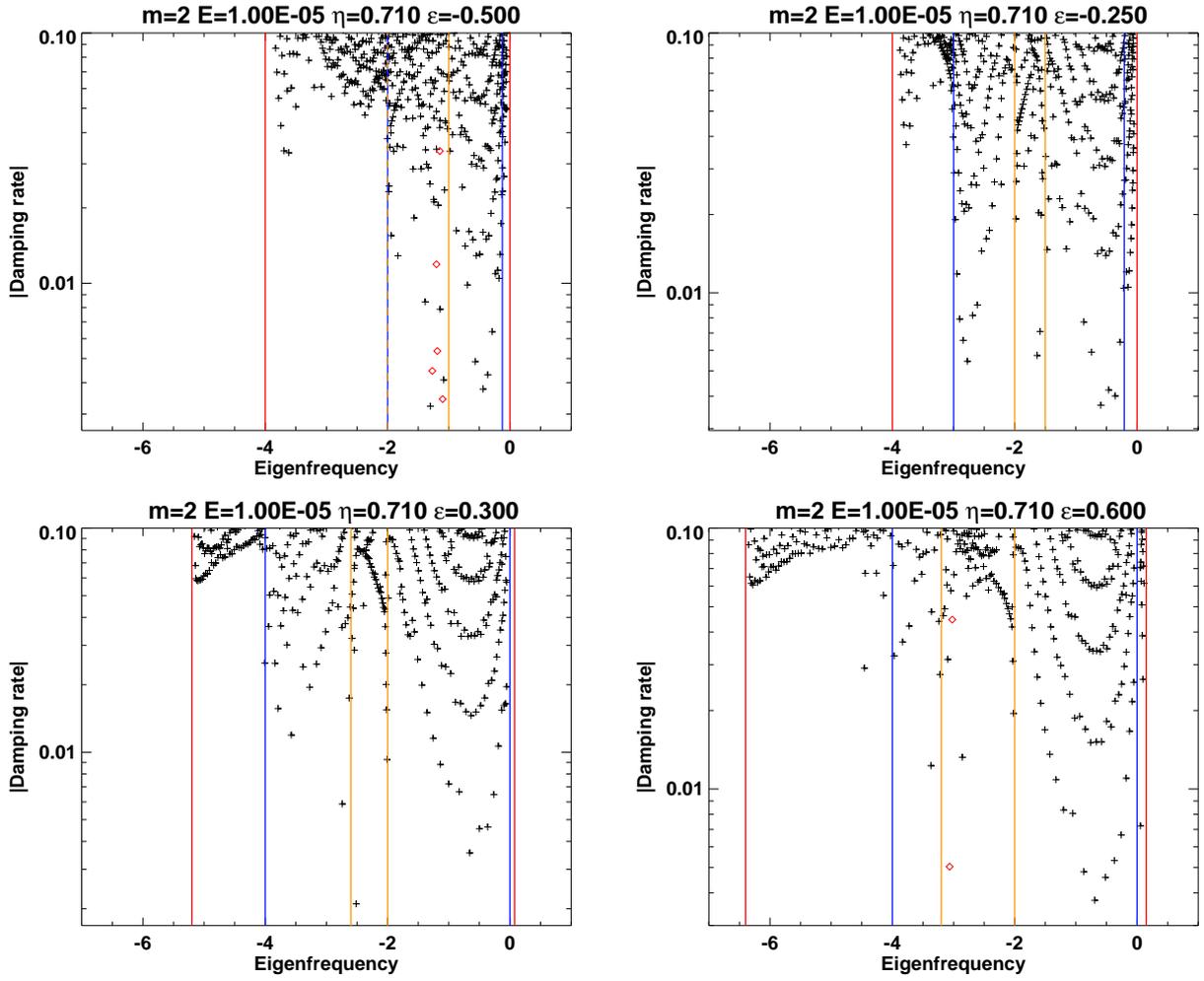

\centering
\includegraphics[width=0.45\textwidth]{{{QZ_m2_eps-0.500_eta0.710_e1.00D-05}}}
\includegraphics[width=0.45\textwidth]{{{QZ_m2_eps-0.250_eta0.710_e1.00D-05}}}
\includegraphics[width=0.45\textwidth]{{{QZ_m2_eps0.300_eta0.710_e1.00D-05}}}
\includegraphics[width=0.45\textwidth]{{{QZ_m2_eps0.600_eta0.710_e1.00D-05}}}
\caption{Same as Fig. \ref{fig:QZ} but for $m=2$ and solar aspect ratio $\eta=0.71$. The four panels are for $\varepsilon=\{-0.50, -0.25, 0.30, 0.60\}$. Orange solid lines are the boundaries of the frequency range in which a corotation layer exists inside the shell (see Fig. \ref{fig:BBR}). Red symbols depict unstable eigenvalues for which ${\rm Re}(i \omega_p) > 0$.}
\label{fig:QZ_3}
\end{figure*}

Our comment on Fig. \ref{fig:QZ} that could explain the rarity of DT modes does not stand in the case of anti-solar differential rotation presented in Fig. \ref{fig:QZ_2}. Though most of the least-damped modes are still D modes, some of them have frequencies that belong to the DT-frequency range, independently of the values of the parameters $\eta$ and $\varepsilon$. This indicates that resonant axisymmetric DT modes in conical differential rotation mostly exist in the anti-solar case ($\varepsilon < 0$), which is consistent with our attempts to compute DT modes, and therefore the QZ factorization method is a precious tool in the exploration of the different families of inertial modes.

As can be seen in Fig. \ref{fig:QZ_3}, the trend indicating that resonant modes preferentially exist in the frequency range of D modes rather than of DT modes is still valid in the case of non-axisymmetric $m=2$ modes. As explained in the introduction we chose $m=2$ because it is the dominant non-axisymmetric component of the tidal potential. The red symbols depict unstable modes which surprisingly exist even at Ekman numbers as high as $10^{-5}$. All of them feature a critical layer inside the shell where the Doppler-shifted wavefrequency $\tilde{\omega}_p$ vanishes, which is reminiscent of the inviscid case exposed at the end of Sect. \ref{sec:inviscid_problem}. We thus expose our results on non-axisymmetric modes with corotation resonances in Sect. \ref{sec:corotation}.

\subsubsection{Examples}

\begin{figure*}
\centering
\includegraphics[width=0.4\textwidth]{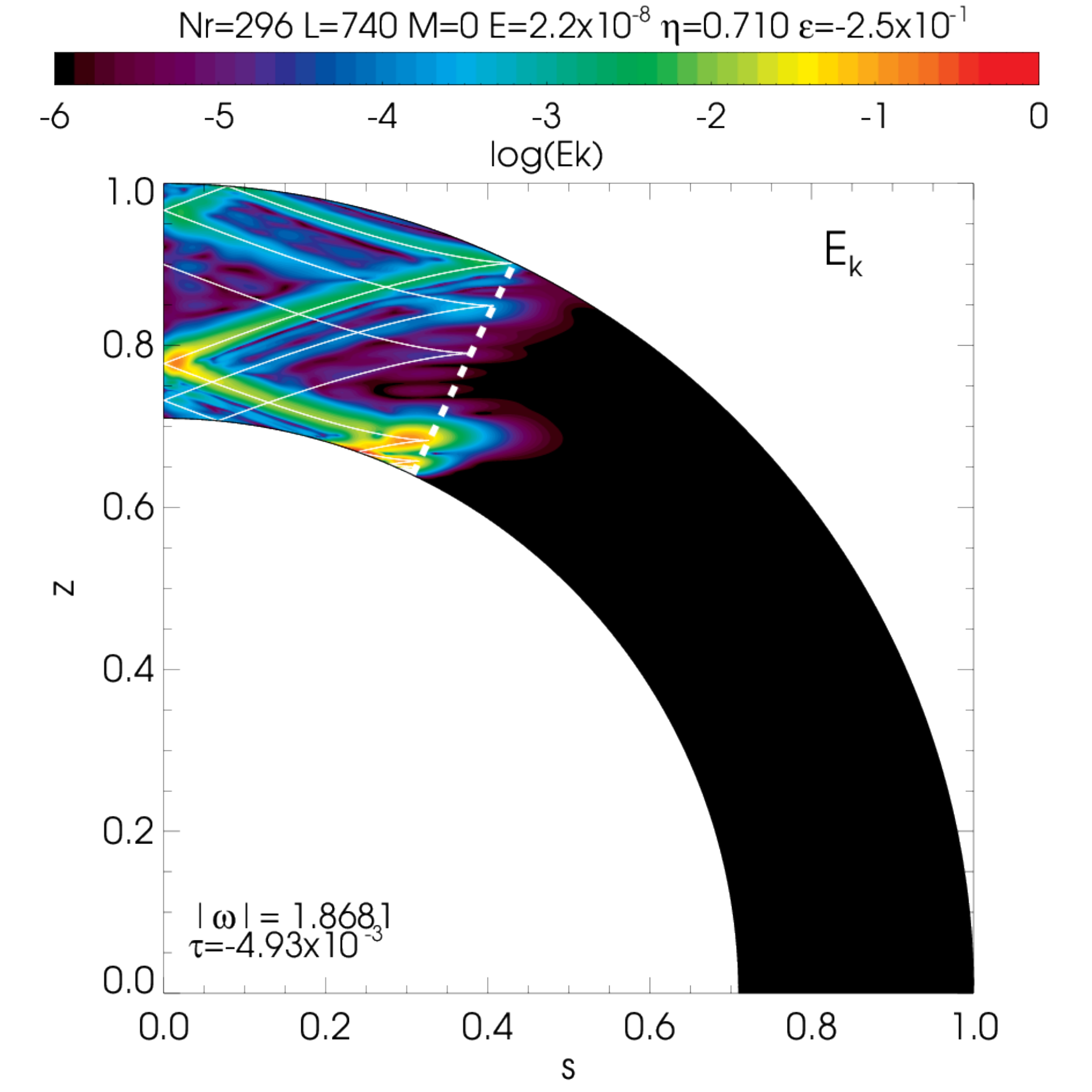} \qquad
\includegraphics[width=0.4\textwidth]{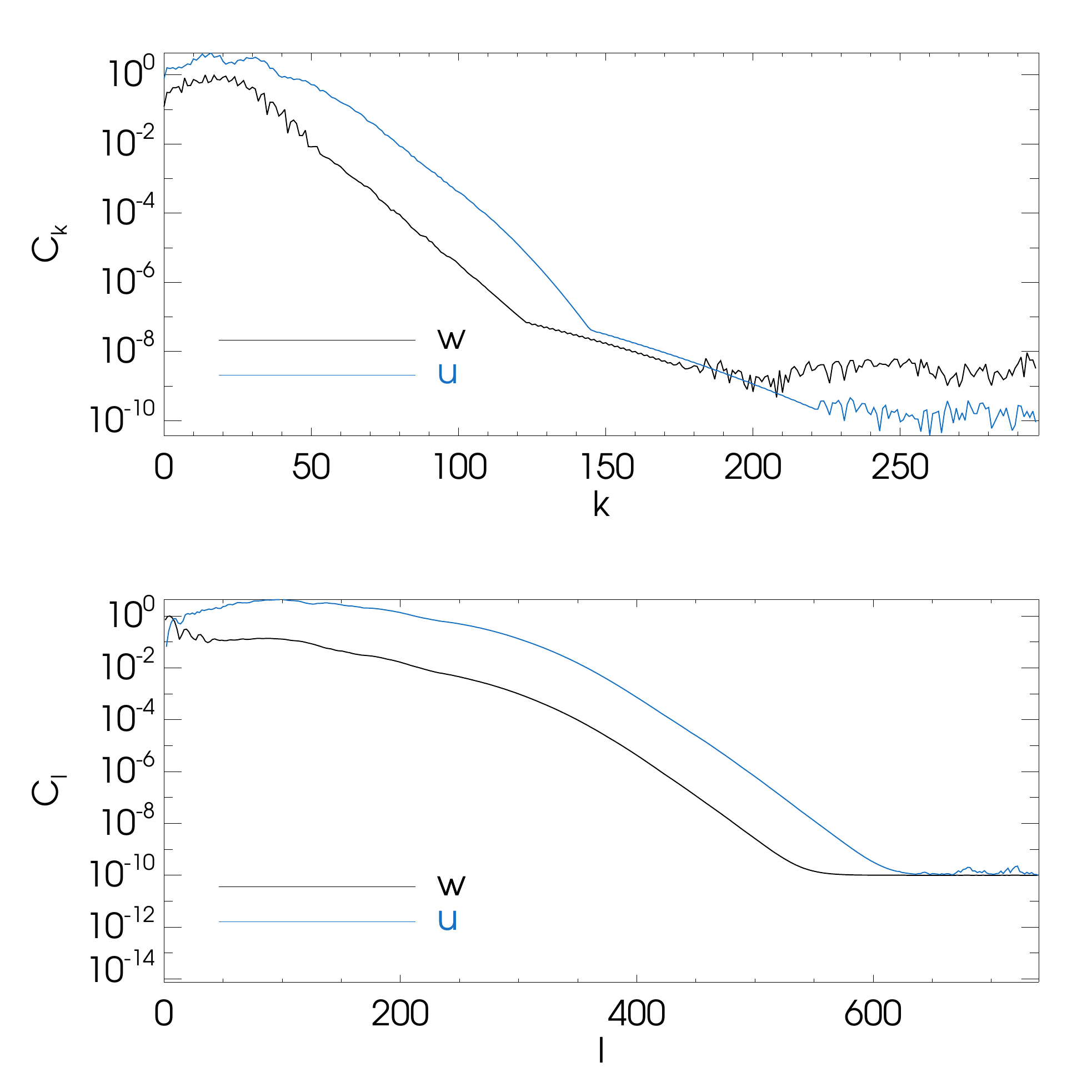}
\caption{{\bf Left~:} Meridional cut of the normalised kinetic energy of a $m=0$ DT mode with eigenfrequency $\omega_p\approx-1.86$ obtained with $E=2.2 \times 10^{-8}$, $\eta=0.71$ and conical differential rotation $\varepsilon=-0.25$. The location of the turning surface is depicted by the white dashed-line. {\bf Right~:} Spectral content of the radial ($u$) and orthoradial ($w$) components of the velocity field for this mode.}
\label{fig:DTmode_0}
\end{figure*}

The axisymmetric DT mode presented in Fig. \ref{fig:DTmode_0} was computed with $\eta=0.71$ and anti-solar rotation parameter $\varepsilon=-0.25$. In order to do this, we first computed a DT mode at eigenfrequency $\omega_p \approx 1.82$ and Ekman number $E=10^{-5}$ before decreasing $E$ step-by-step (typical steps being a relative change of a few percent), using the eigenfrequency of a given step as an initial guess for the next step, along with increasing the spectral resolution. This method gives a better understanding of how the width of the shear layers and/or damping rates scale with $E$ as we approach the astrophysically relevant regime of small $E$. As expected, the mode's kinetic energy is restricted to the hyperbolic domain located near the rotation axis, and the shear layer emitted at the inner critical latitude indeed reflects on the white-dashed line that represents the turning surface. Looking at how paths of characteristics behave for this set of parameters, we find that they eventually focus at the intersection of the inner core with the turning surface, as does the shear layer of the viscous mode.

\begin{figure*}
\centering
\includegraphics[width=0.4\textwidth]{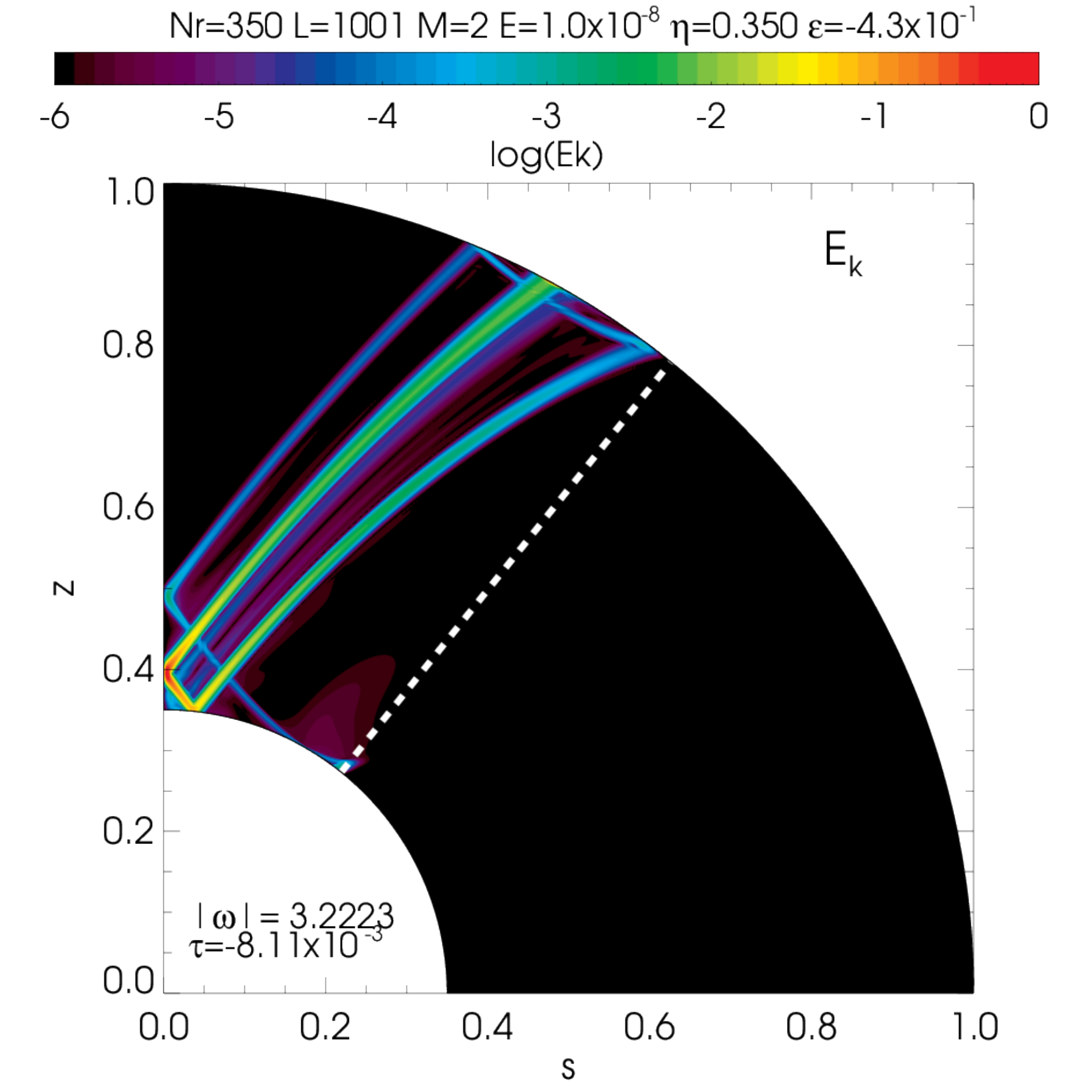}\qquad
\includegraphics[width=0.4\textwidth]{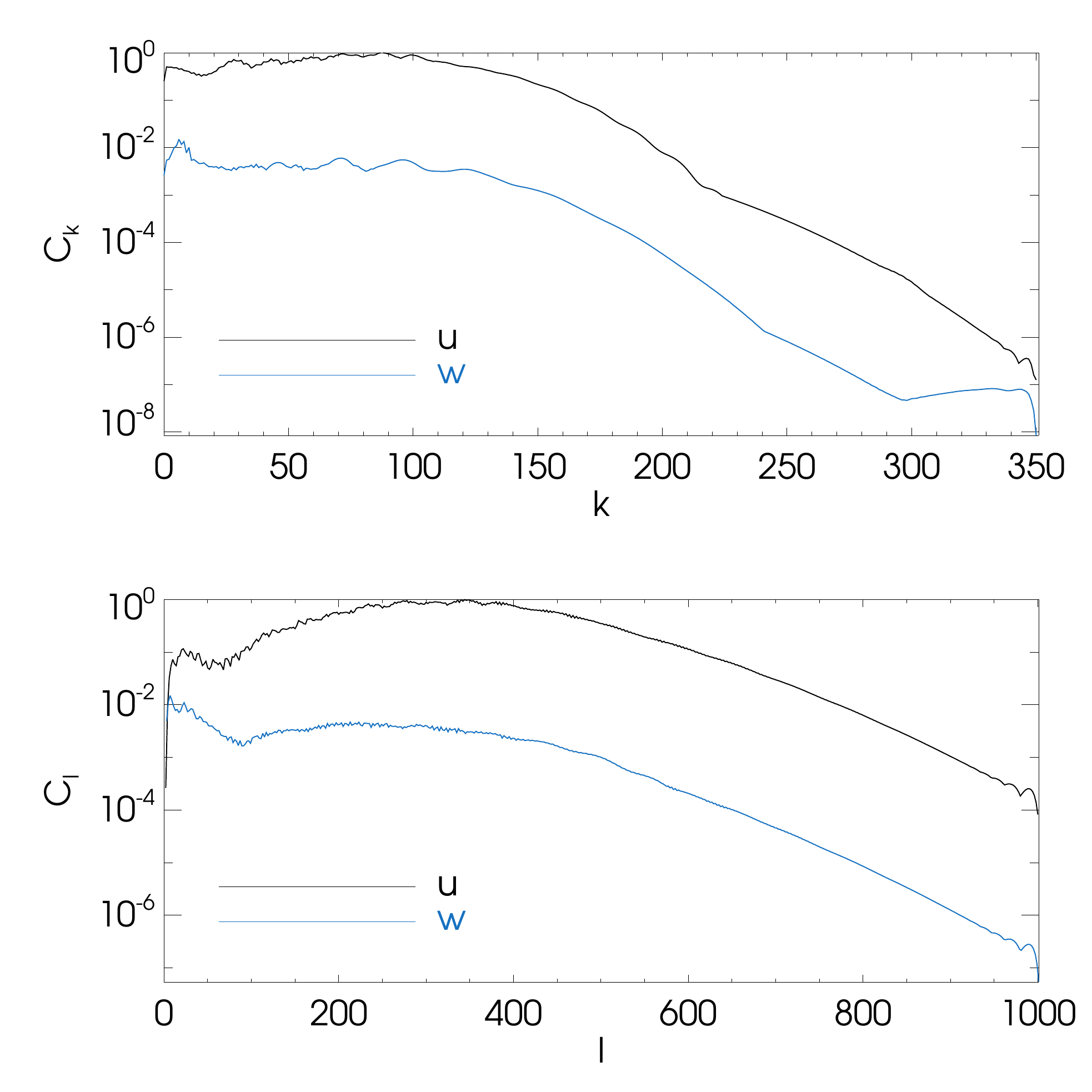}
\caption{{\bf Left~:} Meridional cut of the normalised kinetic energy of a $m=2$ DT mode with eigenfrequency $\omega_p\approx3.22$ obtained with $E=10^{-8}$, $\eta=0.35$ and conical differential rotation $\varepsilon=-0.43$. The location of the turning surface is depicted by the white dashed-line. {\bf Right~:} Spectral content of the radial ($u$) and orthoradial ($w$) components of the velocity field for this mode.}
\label{fig:DTmode_2}
\end{figure*}

Another interesting $m=2$ DT mode is shown in Fig. \ref{fig:DTmode_2} for $\omega_p \approx 3.22$ and $\varepsilon=-0.43$. This mode features a shear layer that passes through both the inner and outer critical latitudes, and which focuses at the intersection between the inner core and the turning surface depicted by the white dashed line in the figure. We carefully checked that for these parameters, paths of characteristics indeed focus at this point.

\subsection{Critical layers}
\label{sec:corotation}

The goal of this section is to show a few results on non-axisymmetric inertial modes for which a corotation layer exists in the shell, which, in the case of our conical rotation profile is --- by construction --- a cone (see Eq. (\ref{eq:corotation})). As pointed out at the end of Sect. \ref{sec:inviscid_problem}, the fact that the Doppler-shifted wavefrequency $\tilde{\omega}_p$ vanishes has important consequences on the propagation properties of inertial waves, since it is a singularity of the inviscid problem. The wave phase velocity ${\bf v_p}$ in the frame rotating with the fluid tends to zero at corotation while the group velocity may either tend to zero, have an infinite vertical component or have a finite slope in the case where $\mathcal{B}=0$.

\begin{figure*}
\centering
\includegraphics[width=0.4\textwidth]{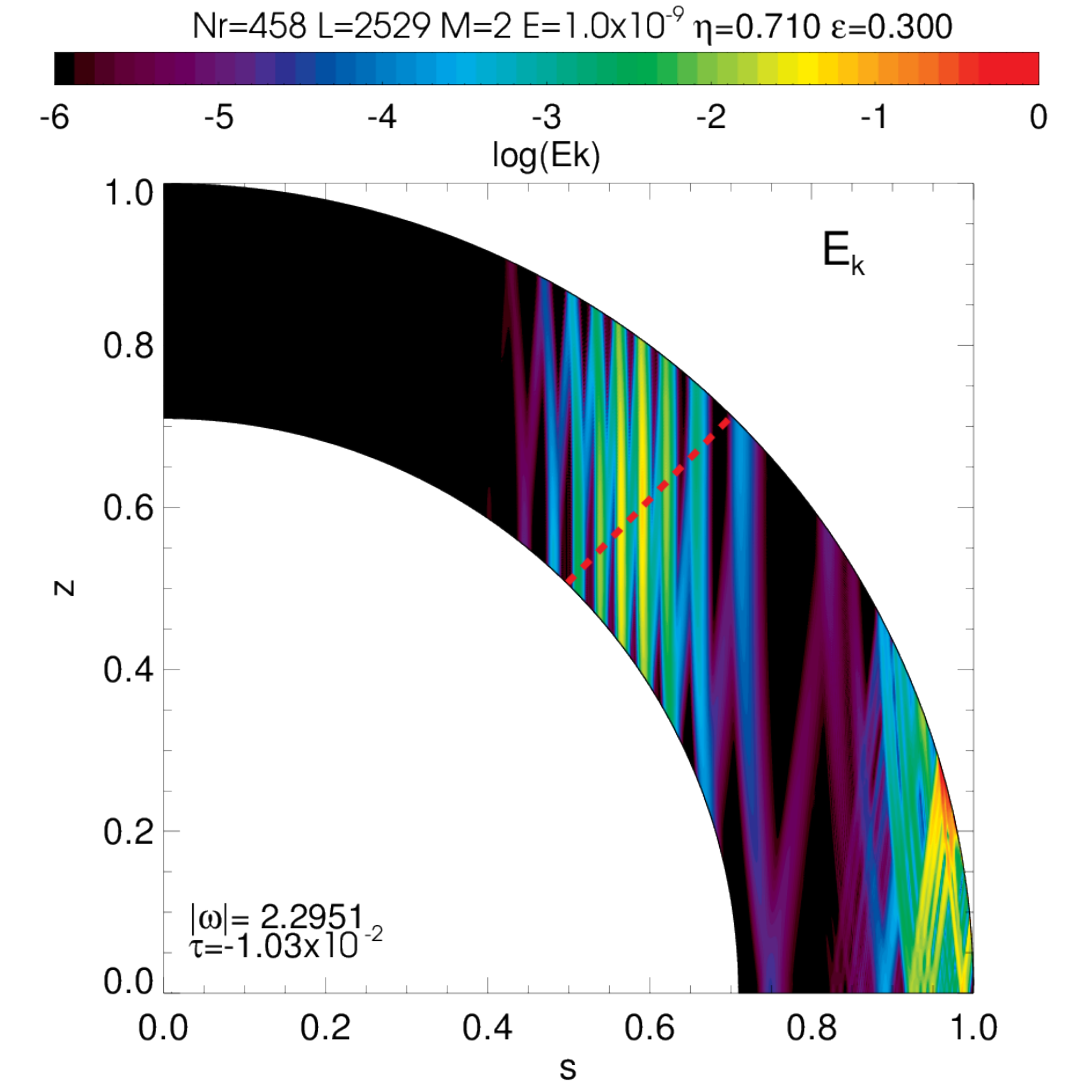}
\includegraphics[width=0.5\textwidth]{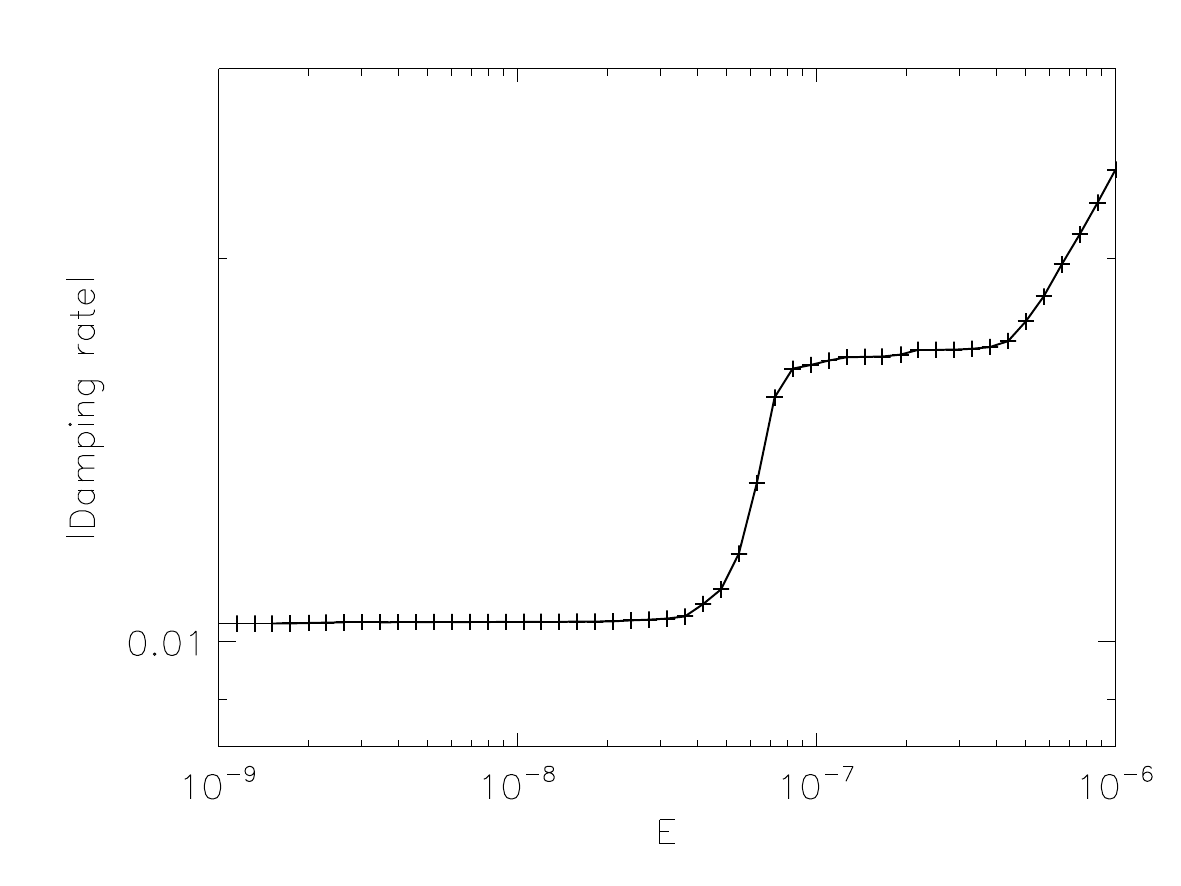}
\caption{{\bf Left~:} Meridional cut of the normalised kinetic energy of an $m=2$ D mode with eigenfrequency $\omega_p\approx-2.30$ obtained with $E=10^{-9}$, the Sun's aspect ratio $\eta=0.71$ and conical differential rotation $\varepsilon=0.3$. The corotation layer is overplotted by the red dashed line. {\bf Right~:} Scaling of the damping rate with respect to the Ekman number $E$. A ``jump'' between two different modes is visible around $E \approx 7 \times 10^{-8}$.}
\label{fig:corotmode_1}
\end{figure*}

Fig. \ref{fig:corotmode_1} features a stable D mode obtained with high spectral resolution for $m=2$ with $\omega_p \approx -2.30$, solar parameters $\eta=0.71$ and $\varepsilon=0.3$ and $E=10^{-9}$. It was obtained by following a mode from $E=10^{-6}$ to $E=10^{-9}$. For these parameters, the corotation resonance is located at $\theta \approx \pi/4$ and is depicted by the red dashed line. A shear layer is emitted at the external critical latitude and the kinetic energy of the mode is maximum there. The shear layer becomes more and more vertical as it approaches the corotation resonance which is in agreement with the propagation properties of characteristics at that location. However, the corotation resonance does not seem to have any other consequence than a local energy accumulation around the critical layer. Unlike the axisymmetric D mode depicted in Fig. \ref{fig:Dmode_2}, the damping rate of this mode does not scale as a power of $E$ but seems to plateau to a constant value. The ``jump'' around $E \approx 7 \times 10^{-8}$ is due to the fact that our method always keeps the least-damped mode at a given step and may therefore switch from one mode to another during the process. Put simply, it might just be an interaction with another eigenvalue.

\begin{figure*}
\centering
\includegraphics[width=0.4\textwidth]{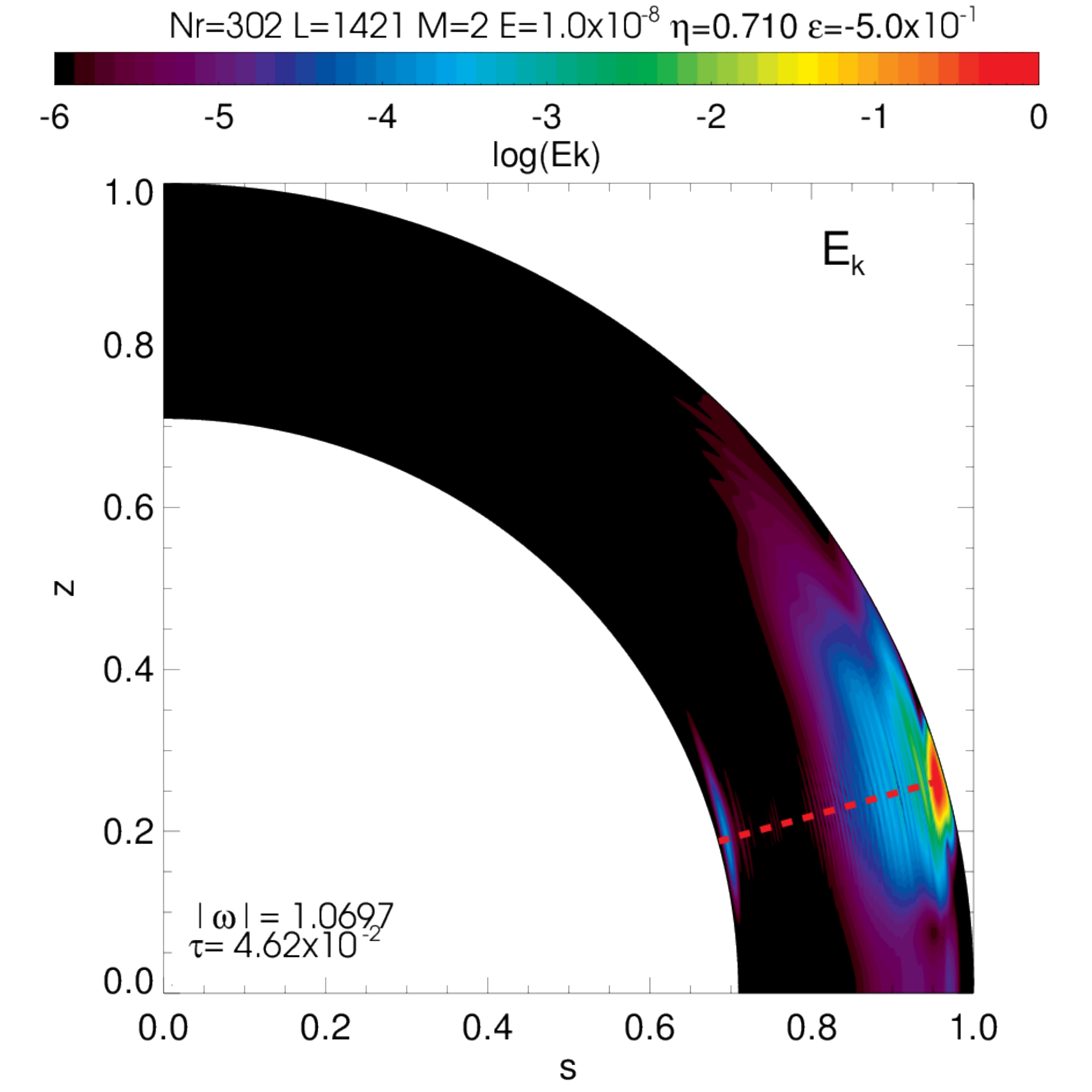}
\includegraphics[width=0.5\textwidth]{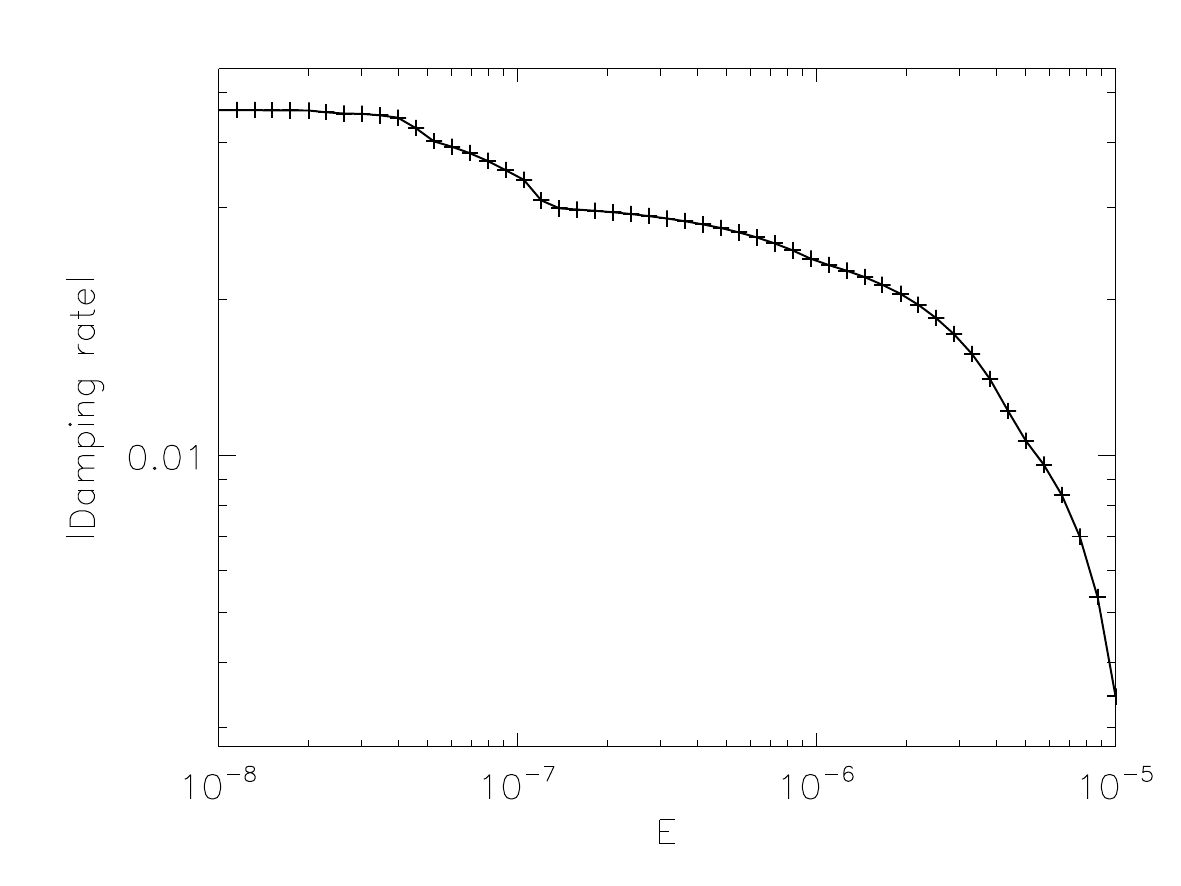}
\caption{{\bf Left~:} Meridional cut of the normalised kinetic energy of an $m=2$ D mode with eigenfrequency $\omega_p\approx-2.30$ obtained with $E=10^{-9}$, the Sun's aspect ratio $\eta=0.71$ and conical differential rotation $\varepsilon=0.3$. The corotation layer is overplotted by the red dashed line. {\bf Right~:} Scaling of the growth rate with respect to the Ekman number $E$. A few ``jumps'' between different modes are visible around $E \approx \left\{5 \times 10^{-6}, 10^{-7}, 5 \times 10^{-8}\right\}$.}
\label{fig:corotmode_2}
\end{figure*}

The case of the mode depicted in Fig. \ref{fig:corotmode_2} is different. We started from the unstable $m=2$ D mode of eigenfrequency $\omega_p \approx -1.10$ along with $\eta=0.71$, $\varepsilon=-0.5$ and $E=10^{-5}$ --- which corresponds to the lower-most red symbol in the top-left panel of Fig. \ref{fig:QZ_3} --- and then progressively decreased $E$ from $10^{-5}$ to $10^{-8}$ with a series of small steps. At each step, we used the eigenfrequency and damping rate of the previous step as the initial guess and selected the least-damped (or most-unstable) mode. The left-panel of Fig. \ref{fig:corotmode_2} displays a cut of the kinetic energy of the unstable velocity field obtained for $E=10^{-8}$. Unlike all the stable oscillation modes presented so far, it shows no recognizable shear layer structure. Instead, a patch of kinetic energy shows up around the location of the critical layer depicted by the red-dashed line. The dependance of the growth rate of this particular mode with Ekman number shown in the right-panel also clearly differs from the classical $E^{1/3}$-scaling presented in Fig. \ref{fig:Dmode_1} for D modes, and is not even a power-law. If we ignore the jumps between different modes (around $E=10^{-7}$ for instance), it seems that for a small enough $E$, the growth-rate becomes roughly constant and reaches a very large value.

The role of critical layers in the propagation properties of different kinds of waves in various containers has been studied before in the fluid mechanics literature : for instance, \cite{Grimshaw1979} studied the effect of critical levels on linear wave propagation as well as the associated absorption \citep{Booker1967} and valve effects \citep{Acheson1972, Acheson1973}. However, these works were restricted to the case where the mean flow is sheared in the vertical direction while our differential rotation profile is sheared in the horizontal direction. On a different approach, \cite{Watson1981} studied the stability of a conical rotation profile similar to ours with respect to non-axisymmetric wavelike perturbations, but was restricted to the case of two-dimensional Rossby waves with negligible radial velocity.

Nevertheless, it seems that critical layers can play a prominent role in the energy and momentum exchanges between the mean background flow and the waves velocity fields and may also strongly impact viscous dissipation. In order to understand the behaviour of inertial waves at corotation resonances, an interesting approach would be the study of a local one-dimensional model in which an inertial wave meets a critical layer induced by a horizontal shear in the background flow. This study is out of the scope of this paper and is postponed to future work.

\section{Conclusions and perspectives}
\label{sec:conclusion}

In this work, we investigated the impact of conical differential rotation on the properties of free inertial waves in a homogeneous fluid inside a spherical shell container. This study is motivated by the possible important role of tidally driven inertial waves in the dynamical evolution of close star-planet or stellar binary systems.

First, we found that differential rotation implies different families of inertial waves and that the frequency range in which they exist can be broadened substantially compared to the solid-body rotation case, provided that a large enough latitudinal differential rotation exists in the convective envelopes of low-mass stars \citep{Brun2002, Brown2008, Matt2011}. We also showed that the D mode regime for which waves can propagate in the entire shell is essentially similar to the solid-body rotation case : wave attractors (which follow curved paths of characteristics) still exist in narrow frequency ranges --- whose number increases with the size of the inner core, see Sect. \ref{sec:inviscid_problem} --- and viscous D modes seem to follow the same scaling relations with Ekman number. Moreover, the least-damped modes seem to be found preferentially in the D mode frequency range. In addition, differential rotation also affects their allowed propagation domain : in the DT regime, modes are trapped around the poles or the equator by a turning surface. It seems however that resonant DT modes are rare especially for solar-like differential rotation. It is important to note that these conclusions only stand for axisymmetric modes or non-axisymmetric modes without corotation resonances. Indeed, we found that these corotation layers strongly modify the behavior of inertial waves since the wave phase velocity always vanishes there, and the paths of characteristics and the associated viscous shear layers may become vertical while crossing the resonance. Among all the numerical calculations performed in the viscous case, we mostly found stable modes, but we also found several unstable modes at Ekman numbers as high as $10^{-5}$ which all featured a corotation resonance. As discussed when introducing the physics of the model, these unstable modes delineate the range of allowed parameters, since the background flow needs to be stable.

From the astrophysical perspective, these conclusions could have important dynamical implications for systems where inertial waves are excited, sustained and finally dissipated in a low-mass star with latitudinal gradient of differential rotation in its convective envelope. One possible source of this excitation is tides raised by a close-in stellar or planetary companion. This requires that one (or more) of the components of the tidal potential --- which depends on the orbital configuration of the system --- has a frequency that falls into the inertial range, with a high enough amplitude so that the kinetic energy stored by inertial oscillations and the subsequent viscous dissipation alters the orbital and rotational dynamics of the system. In a subsequent paper, we shall derive analytically the forcing term due to tides before computing numerically the velocity field and viscous dissipation of such tidally-excited inertial waves. This will allow us to investigate the influence of differential rotation --- e.g. through turning surfaces, corotation layers --- in the convective envelope of low-mass stars  on tidal dissipation for various parameters related to stellar mass, rotation, companion mass and orbital parameters.

Finally, we point out the limitations and possible improvements of our work. Beside the conditions of stability that we already mentioned, we may note that our model only accounts for the direct effects of the viscous dissipation of inertial waves in convective envelopes, which may not be the only mechanism responsible for tidal dissipation --- the non-linear breaking of gravity waves in the radiative interior is another mechanism, see \cite{Barker2010, BarkerOgilvie2011, Barker2011} ---, especially in cases where these waves are likely not to be excited. Additionally, we point out that our study assumes the fluid to be incompressible and does not include the effects of non-linearities \citep[see][]{Jouve2014, Favier2014} which could become important for tidal oscillations of high amplitude, or other processes which could affect the very nature of the low-frequency oscillations studied here --- such as magnetic field, centrifugal distortion \citep{Braviner2014, Barker2014} or the possible interactions with turbulent convective motions \citep{Ogilvie2012}. However the effects of differential rotation on the physical processes driving tidal dissipation have not been investigated so far, making this work a step towards a more detailed understanding of binary systems and star-planet interactions.

\begin{acknowledgements}
We warmly thank the referee A. J. Barker for his constructive comments which allowed us to improve the paper. M. Guenel and S. Mathis acknowledge funding by the European Research Council through ERC grant SPIRE 647383. This work was also supported by the Programme National de Plan\'etologie (CNRS/INSU) and CNES PLATO grant at CEA-Saclay.
\end{acknowledgements}

%=========%=========%=========%=========%=========%=========%=========%=
%BIBLIOGRAPHIE
%=========%=========%=========%=========%=========%=========%=========%=
\bibliographystyle{aa}  % A&A bibliography style file (aa.bst)
\bibliography{GBMR2015} % references .bib file

\end{document}